\documentclass[lettersize,journal]{IEEEtran}

\usepackage{cite}
\usepackage{amsmath,amssymb,amsfonts}
\usepackage{graphicx}
\usepackage{textcomp}
\usepackage[table,xcdraw]{xcolor}
\usepackage[flushleft]{threeparttable}
\usepackage{subfig}
\usepackage{algorithm}
\usepackage[noEnd=true]{algpseudocodex}
\usepackage{booktabs}
\usepackage{url} 
\usepackage[export]{adjustbox}
\usepackage{siunitx}
\usepackage{orcidlink}
\usepackage{dblfloatfix}
\usepackage{framed}

\usepackage[dvipsnames]{xcolor}

\usepackage{tikz}
\usepackage{tikzsymbols}
\usetikzlibrary{spy, positioning, shapes, shapes.geometric, arrows, chains, patterns, decorations.markings, pgfplots.groupplots}

\usepackage{pgfplots,pgfplotstable} 
\usepgfplotslibrary{statistics}

\definecolor{mycolor1}{RGB}{251,180,174}
\definecolor{mycolor2}{RGB}{179,205,227}
\definecolor{mycolor3}{RGB}{204,235,197}
\definecolor{mycolor4}{RGB}{222,203,228}
\definecolor{mycolor5}{RGB}{254,217,166}
\definecolor{mycolor6}{RGB}{255,255,204}
\definecolor{mycolor7}{RGB}{8,29,88}
\definecolor{wildstrawberry}{rgb}{1.0, 0.26, 0.64}

\pgfplotstableset{
  create on use/tiratio/.style={
      create col/expr={\thisrow{tistrat}*\thisrow{imp}}
  }
  create on use/fspfiratio/.style={
      create col/expr={\thisrow{fspstrat}*\thisrow{imp}}
  }
}

\def\BibTeX{{\rm B\kern-.05em{\sc i\kern-.025em b}\kern-.08em
    T\kern-.1667em\lower.7ex\hbox{E}\kern-.125emX}}
\begin{document}

\title{A Zero-overhead Flow for Security Closure}

% \author{
% \IEEEauthorblockN{
% Mohammad Eslami\textsuperscript{1}, 
% Ashira Johara\textsuperscript{2}, 
% Kyungbin Park\textsuperscript{2}, 
% Samuel Pagliarini\textsuperscript{2,1}}
% \IEEEauthorblockA{
% \textsuperscript{1}\textit{Tallinn University of Technology}, %Tallinn, Estonia \\ 
% \textsuperscript{2}\textit{Carnegie Mellon University}, Pittsburgh, %PA, USA \\
% Email: mohammad.eslami@taltech.ee, \{ajohara, kyungbip\}@andrew.cmu.edu, pagliarini@cmu.edu
% }
% }

\author{Mohammad Eslami\orcidlink{0000-0001-7200-3655}, 
Ashira Johara,
Kyungbin Park\orcidlink{0009-0007-8887-6128},
and Samuel Pagliarini\orcidlink{0000-0002-5294-0606},~\IEEEmembership{Member,~IEEE}% <-this % stops a space
\thanks{This work was partially supported by the EU through the European Social Fund in the context of the project “ICT programme”.}
\thanks{M. Eslami is with the Department of Computer Systems, Tallinn University of Technology (TalTech), 12618, Tallinn, Estonia (e-mail: mohammad.eslami@taltech.ee)}
\thanks{Ashira Johara and Kyungbin Park are with the ECE Department, Carnegie Mellon University, 15213, Pittsburgh, PA, USA (e-mail: \{ajohara, kyungbip\}@andrew.cmu.edu)}
\thanks{Samuel Pagliarini is with the Department of Computer Systems, Tallinn University of Technology (TalTech), 12618 Tallinn, Estonia, and also with the ECE Department, Carnegie Mellon University, 15213, Pittsburgh, PA, USA (e-mail: pagliarini@cmu.edu).}}

% \and
% \and
% \and
% \and
% \and

\maketitle

\begin{abstract}
In the traditional Application-Specific Integrated Circuit (ASIC) design flow, the concept of timing closure implies to reach convergence during physical synthesis such that, under a given area and power budget, the design works at the targeted frequency. However, security has been largely neglected when evaluating the Quality of Results (QoR) from physical synthesis. In general, commercial place \& route tools do not \emph{understand} security goals. In this work, we propose a modified ASIC design flow that is security-aware and, differently from prior research, does not degrade QoR for the sake of security improvement. Therefore, we propose a first-of-its-kind zero-overhead flow for security closure. Our flow is concerned with two distinct threat models: (i) insertion of Hardware Trojans (HTs) and (ii) physical probing/fault injection. Importantly, the flow is entirely executed within a commercial place \& route engine and is scalable. In several metrics, our security-aware flow achieves the best-known results for the ISPD'22 set of benchmark circuits while incurring negligible design overheads due to security-related strategies. Finally, we open source the entire methodology (as a set of scripts) and also share the protected circuits (as design databases) for the benefit of the hardware security community.
\end{abstract}

\begin{IEEEkeywords}
ASIC flow, security closure, place and route, physical synthesis, hardware security
\end{IEEEkeywords}

\section{Introduction}

As Integrated Circuits (ICs) become foundational to modern electronics, their design and manufacturing processes are increasingly complex and often outsourced, introducing potential security vulnerabilities~\cite{HwSOverview}. Traditionally, Application Specific Integrated Circuit (ASIC) design flows have prioritized Power, Performance, and Area (PPA) optimizations, but this focus is insufficient in the face of evolving security threats. Yet, no commercial Place and Route (P\&R) engine is security-aware. Several recent academic works discuss security awareness with respect to physical synthesis~\cite{JohannICCAD, Security_Closure, Security_Closure2, contest, eslami23}, including two large-scale blue team versus red team contests~\cite{contest,eslami23,chescontest}.

%Unlike PPA metrics, which can be adjusted and optimized in the later design stages, security measures embedded at the physical layout level are nearly impossible to implement retrospectively. As a result, security closure is quickly becoming essential in the ASIC design process to ensure resilience against malicious threats introduced both during and after fabrication. 

The layout of an IC, which serve as the blueprint for chip fabrication, is vulnerable to a range of sophisticated attacks~\cite{salmani, rethinking, tvf, hwSecurity}. Among those, Hardware Trojans (HTs) are small but potentially devastating malicious modifications that can be introduced at various points in the supply chain, including at fabrication time~\cite{TrjClass, ProtectTrj, TrjLessons, ECOtrojan, HT_1,HT_2}. Some threats take place after ICs are fabricated, particularly in the form of probing~\cite{probing, probing_2} and Fault Injection (FI)~\cite{FT_father_2, FT_father}. Here, the objective of the adversary is either to readout some sensitive/privileged information or to corrupt it. Sophisticated equipment can enable adversaries to access specific IC layers through Front-Side Probing (FSP)~\cite{tehraniFSP}, while back-side probing offers an alternative path to reading or injecting values \emph{almost} directly on transistors~\cite{bs_probing, backside, backside_2}. FI attacks rely on laser pulses, electromagnetic interferences, or voltage manipulation to disrupt IC functionality~\cite{ChipFI2, chipFI, laserFI}. FI attacks can reveal sensitive data or cause operational failures that undermine the integrity of a system~\cite{FI_1, FI_2}. 

The growing sophistication of attacks highlights the need for integrated, proactive security measures within ASIC design flows. Given the difficulty of retrofitting security into hardware after fabrication, design-time security closure is essential to counteract vulnerabilities. \textbf{Security closure}~\cite{JohannICCAD}, i.e., to address design-time security considerations, has emerged as a concept that is the counterpart to conventional timing closure. Unlike optimizing solely for traditional QoR metrics, security closure entails optimizing the design for resilience against specific adversarial techniques and corresponding threat models. Several methodologies that incorporate elements of security closure have been proposed~\cite{Security_Closure, Security_Closure2, contest, ASSURER, gdsguard, JohannICCAD, TroLLoc, TroMUX, salsy, DEFend}. In general, the proposed schemes promote changes to the place and route (P\&R) solutions such that security-critical cells or wires are repositioned. Many security-focused layout modifications introduce overheads, affecting the IC's performance, power consumption, and/or area \cite{TroLLoc, TroMUX, JohannICCAD, gdsguard, salsy}.

A key challenge for the adoption of security closure, without a doubt, is its associated overheads. In this paper, we propose a \textbf{security-aware} ASIC design flow that is seamlessly integrated with commercial physical synthesis tools. Unlike prior approaches that trade PPA for protection against HTs and/or FSP/FI, we preserve and prioritize PPA. The main contributions of this paper are:
\begin{itemize}
    \item \emph{No compromise on PPA}: Our approach achieves security closure while maintaining PPA targets.
    \item \emph{Open source release}: all scripts and design databases associated with our methodology are open sourced \cite{github}.
    %\footnote{To preserve the blind review process, all files are already available in an anonymous repository: \url{https://anonymous.4open.science/r/zero-overhead/}}.%, this promoting transparency, reproducibility, and adoption. 
    \item \emph{Scalability and compatibility}: By utilizing commercial P\&R tools, our security-aware design flow is highly scalable and compatible with existing industry workflows.%, supporting a broad range of ASIC projects.
\end{itemize}

The rest of this paper is organized as follows: Section~\ref{sec:bg} provides a comprehensive background of related works. In Section~\ref{sec:flow}, we detail the zero-overhead flow proposed for security closure and its implications. Section~\ref{sec:results} presents our experimental results. A discussion is provided in Section~\ref{sec_discusstion}. Finally, Section~\ref{sec:conclusion} concludes the paper.

\section{Background} \label{sec:bg}

\subsection{International Symposium on Physical Design Contest}

\begin{figure*}[b]
\begin{equation}{\label{eq1}}
    Score = Design Quality \times Security = DES \times \frac{(TI+FSPFI)}{2}
\end{equation}
\end{figure*}

\begin{figure*}[b]
\begin{equation} \label{eq2}
\begin{split}
Score = \overbrace{\left(0.1 \times \frac{(des\_p\_total)_{\text{sec}}}{(des\_p\_total)_{\text{bl}}} + 0.3 \times \frac{(des\_perf)_{\text{sec}}}{(des\_perf)_{\text{bl}}} + 0.3 \times \frac{(des\_area)_{\text{sec}}}{(des\_area)_{\text{bl}}}+ 0.3 \times \frac{(des\_issues)_{\text{sec}}}{(des\_issues)_{\text{bl}}}\right)}^{\text{\emph{DesignQuality (DES)}}} \\ 
\times \left( \overbrace{ \frac{1}{2} \times \left(0.5 \times \frac{(fsp\_fi\_ea\_c)_{\text{sec}}}{(fsp\_fi\_ea\_c)_{\text{bl}}} + 0.5 \times \frac{(fsp\_fi\_ea\_n)_{\text{sec}}}{(fsp\_fi\_ea\_n)_{\text{bl}}}\right) }^{\text{\emph{Security (FSPFI)}}}
+ \overbrace{ \frac{1}{2} \times \left(0.6 \times \frac{(ti\_sts)_{\text{sec}}}{(ti\_sts)_{\text{bl}}} + 0.4 \times \frac{(ti\_fts)_{\text{sec}}}{(ti\_fts)_{\text{bl}}}\right)}^{\text{\emph{Security (TI)}}} \right) 
\end{split}
\end{equation}
\label{fig:your_label}
\end{figure*}

The ISPD 2022 contest was titled ``Security closure of physical layouts'' and it aimed to enhance the security of digital IC layouts during physical synthesis against various hardware security threats. Contest participants, acting as the defenders, were challenged to implement physical design measures to protect twelve different designs against three major threats:
\begin{itemize}
    \item HT insertion: Preventing the addition of malicious logic during fabrication.
    \item Probing attacks: Protecting the IC's frontside from attacks that attempt to readout data from the wires.
    \item Fault injection: Protecting the IC from the attacker who tries to induce faults onto the IC.
\end{itemize}

The contest evaluated both the \textbf{security} and \textbf{design quality} of the submitted layouts using a combined, weighted scoring formula, as summarized in Eq.~\ref{eq1} and expanded in Eq.~\ref{eq2}.

The Design Quality metric (DES) summarizes the performance characteristics of the design (power, area, timing), while the Security metric assesses the effectiveness of implemented security measures, focusing on Trojan Insertion (TI) and Frontside Probing and Fault Injection (FSPFI) combined.

Equation~\ref{eq2} presents a detailed breakdown of the simplified formula given in Eq.~\ref{eq1}, where \emph{DesignQuality} is expressed as the weighted sum of four factors: power consumption, measured as total power (\emph{des\_p\_total}); performance, reflected in timing behavior such as worst and total negative slack (\emph{des\_perf}); area, quantified as total die area (\emph{des\_area}); and routing quality, indicated by the number of Design Rule Check (DRC) violations (\emph{des\_issues}). It should be noted that the metrics are normalized against baseline layouts provided by the contest organizers.

%This metric encompasses several key aspects: power consumption, measured as total power; performance, measured using the worst and total negative slack for setup timing requirements; area, measured as total die area; and routing, measured by the number of Design Rule Check (DRC) violations. More details are available from~\cite{contest}.

The Security components in Eq.~\ref{eq2} complement the Design Quality component by quantitatively assessing the design's robustness against FSPFI and TI threats. It is computed as the average of two equally weighted components, each targeting a specific attack vector. TI is evaluated based on the notion of ``vulnerable regions,'' which are areas in the layout where an attacker could potentially insert an HT. The scoring considers the number and the size of these regions, as well as the availability of free routing tracks for connecting the HT to the original circuitry. In Eq.~\ref{eq2}, \emph{ti\_sts} and \emph{ti\_fts} denote the number of exploitable placement sites and the available routing tracks around those regions.

FSPFI is evaluated based on the notion of ``exposed area'' of sensitive cells and nets, which refers to the portion of these components that are directly accessible through the metal stack (from the front side). The score considers the total, maximum, and average exposed area for both cells and nets. In Eq.~\ref{eq2}, the terms (\emph{fsp\_fi\_ea\_c}) and (\emph{fsp\_fi\_ea\_n}) correspond to the exposed area of the cell assets and net assets, respectively. For each design, organizers provided a subset of cells and nets declared as \emph{assets}, indicating that they merit protection against FSPFI.

The final score is calculated as the product of the weighted design quality component and the weighted security component, thus ensuring a good score reflects not only a secure layout but also one that generally maintains good performance. As previously mentioned, the metrics in Eq.~\ref{eq1} are normalized with respect to baseline layouts: A score of 1 represents no change from the baseline, a score below 1 indicates improvement, and a score above 1 implies deterioration. 

% \subsection{Improved Scoring formula}
While the scoring methodology adopted in the ISPD'22 contest aimed to balance security and design quality, it allowed for DRC violations. Even if these violations were penalized by the design quality metric, this approach does not align with real-world chip implementation practices where absolutely no DRCs are allowed. In our own evaluation, we adopt the same scoring formula but enforce \textbf{zero DRC compliance} as a hard constraint. In other words, we operate under a harsher scoring system than that introduced in~\cite{contest}.

%The most significant concern with the scoring formula is the zero-product issue. Since the overall score is the product of the Design Quality score and the Security score. This means that if a layout archives a score of 0 in either category, the overall score becomes 0, rendering the other category meaningless. For instance, a layout with exceptional security measures but poor design quality (or vice versa) would receive an overall score of 0, failing to reflect the strength of layout in the non-zero category. This can discourage the participants from innovating in one category and if they perceive that achieving a high score in the other category is unattainable. 

%Participants eventually figured out that they could achieve a score of 0 for the security side of the scoring formula, which is composed of Trojan Insertion (TI) and Frontside Probing/Fault Injection (FSPFI) metrics. To achieve this, they had to make each of the security metrics 0. For TI, the design should not have any exploitable regions. For the FSPFI, if participants placed a large metal layer that covers the total surface of the chip, it would hide the whole exposed area of the sensitive parts, resulting in FSPFI metrics of 0. Consequently, the total Security metric becomes 0, and participants no longer had to care about the Design Quality metric since the overall score was already 0. This loophole undermines the contest's objective of balancing security and design quality, as it allows participants to disregard design quality entirely. 

\subsection{Related Works}

Several studies have addressed the challenge of security closure in physical layouts and its impact on PPA. %In \cite{JohannICCAD}, the authors discuss the challenges and strategies involved in the concept of security closure at the physical layout level. This paper introduces DEFence, which is a flexible CAD framework at the layout level, designed to evaluate and address the post-design threats throughout the physical design process. Moreover, it emphasizes the importance of integrating security closure within existing commercial CAD flows. However, user control over placement may be constrained in this approach, especially when relying on commercial CAD tools.   
In~\cite{JohannICCAD}, authors introduced DEFence, a flexible CAD framework for addressing post-design threats and integrating security closure at the physical layout level. While this work aims to minimize PPA impact, it does not provide concrete data or analysis to demonstrate the actual overhead incurred. %However, user control over placement may be limited when using commercial CAD tools. 
In \cite{TroLLoc}, authors proposed TroLLoc, a scheme combining logic locking and layout hardening to prevent Trojan insertion. A similar approach based on logic locking is presented in~\cite{TroMUX}, where the authors present TroMUX, a Mux-based logic locking scheme integrated into physical synthesis to secure critical components and minimize open placement sites. As is the case with most logic locking solutions, TroLLoc and TroMUX require the use of a tamper-proof memory that is not trivial to obtain. Additionally, instantiating such memory adds significant cost and complexity, including floorplanning decisions. %while remaining vulnerable to advanced attacks.
ASSURER, a framework for security closure with reduced PPA impact, is introduced in~\cite{ASSURER}. ASSURER uses a reward-directed refinement and multi-threshold partitioning to prevent HT insertion. The framework also includes a probing attack prevention flow based on Engineering Change Order (ECO) routing. However, this approach is vulnerable to attacks reversing the refinement process to create zones for malicious logic insertion. In~\cite{gdsguard}, the authors propose GDSII-Guard, which too uses ECO features and a multi-objective optimization model to enhance layout-level security while balancing PPA. However, the approach often degrades timing, leading to negative timing slack. Additionally, since it proposes customized solutions for different designs by random exploration of parameters, the scalability and practicality of the approach are uncertain. 

%In \cite{gdsguard}, the authors propose GDSII-Guard, which uses ECO features to optimize security at the layout level. GDSII-Guard employs a multi-objective optimization model to balance security and PPA. The approach focuses on minimizing the number and the size of exploitable regions to make the HT insertion more challenging. This technique, however, generally degrades timing, often resulting in negative timing slack within the design.

%SALSy, which is a design-time methodology for protecting ICs against fabrication-time and post-fabrication attacks, is introduced in~\cite{salsy}. It diverges from other works by focusing on techniques applied during the physical synthesis and validating its effectiveness through an actual silicon demonstration using commercial technology. SALSy demonstrates the feasibility of implementing security closure techniques in a real-world setting. Nonetheless, the reliance on buffer insertion for filling the open placement sites, while effective, also leads to increased power consumption.
In \cite{salsy}, we have introduced SALSy, a design-time methodology for securing ICs against fabrication and post-fabrication attacks. SALSy stands out as the first security closure approach validated in silicon, but its reliance on buffer insertion for populating regions vulnerable to TI increases power consumption. The same work also provides a thorough discussion on academic PDKs versus commercial PDKs from a security closure point of view.

In general, it can be argued that most existing solutions prioritize security at the cost of PPA, which limits their practical applicability. In some cases, a co-optimization between PPA and security is carried out, which also brings PPA overheads, even if they are deemed acceptable. This emphasizes the need for innovative methods that can provide robust security without compromising PPA. Our proposed methodology addresses this critical need. 

\section{Zero-overhead Flow for Security Closure} \label{sec:flow}

Our methodology for security closure is divided in three stages: implementation strategy (IMP), TI strategy, FSPFI strategy. The stages are executed in order, one after another, a single time each. The entire flow is developed as TCL scripts and is executed within Cadence Innovus. Let us start by discussing our implementation strategy.

\subsection{IMP strategy}

Before applying any security-specific approaches to our designs, we focus on minimizing power and area while meeting timing. We highlight that we utilize an \emph{industry-grade} flow that is significantly more intricate than a single pass textbook P\&R flow. Our flow alternates optimization targets (i.e., it switches between timing and power targets) while also setting margins for setup timing dynamically. Moreover, we adhere to the same strict rules of the ISPD contest with respect to power meshes and pin locations for IOs, which are kept identical to the baseline designs provided by the organizers. The entire flow is depicted in Fig.~\ref{fig:flow}.

\definecolor{OliveGreen}{RGB}{ 69, 108, 74 }
\definecolor{RoyalPurple}{RGB}{ 24, 57, 123  }
\definecolor{ProcessBlue}{RGB}{ 50, 160, 245 }
\definecolor{RoyalBlue}{RGB}{ 24, 57, 123 }
\definecolor{ForestGreen}{RGB}{ 10, 130, 37 }

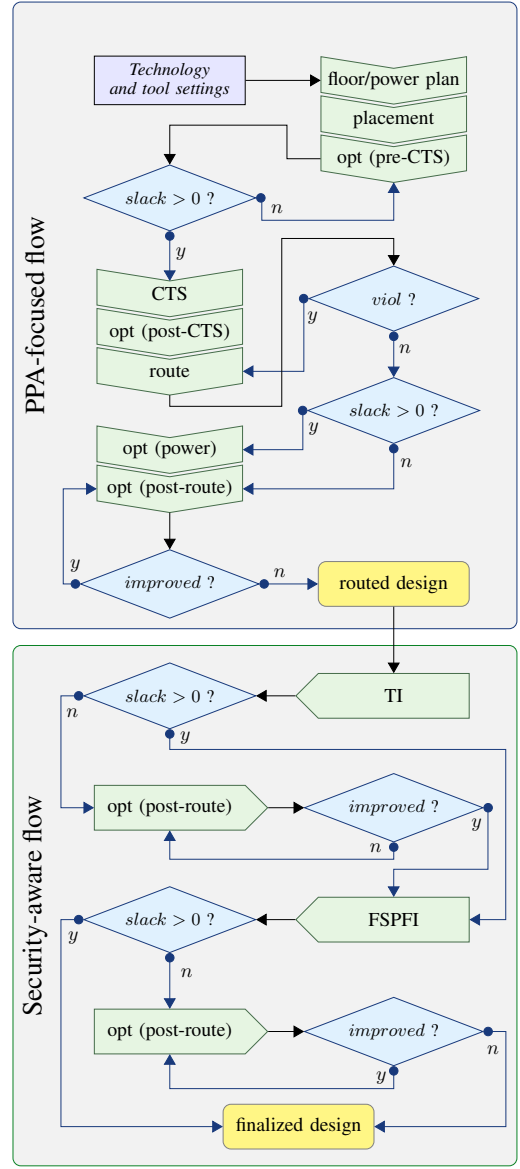
\begin{figure}
\centering
\begin{adjustbox}{width=0.37\textwidth}
\begin{tikzpicture}[%
    >=triangle 60,              % Nice arrows; your taste may be different
    start chain=going below,    % General flow is top-to-bottom
    node distance=5mm and 60mm, % Global setup of box spacing
    every join/.style={norm},   % Default linetype for connecting boxes
    scale=0.2
    ]

    \tikzset{
  base/.style={draw, on chain, on grid, align=center, minimum height=3ex},
  proc/.style={base, rectangle, text width=7em, minimum height=8mm, fill=mycolor3!50, draw=OliveGreen},
  fproc/.style={base, signal, text width=7em, signal from=west, yscale=.16, minimum height=2.6cm, rotate=270, fill=mycolor3!50, draw=OliveGreen},
  rproc/.style={base, signal, text width=7em, minimum height=.8cm, fill=mycolor3!50, rotate=180, draw=OliveGreen},
  lproc/.style={base, signal, text width=7em, minimum height=.8cm, fill=mycolor3!50, draw=OliveGreen},
  longproc/.style={base, rectangle, text width=7em, fill=blue!10},
  test/.style={base, diamond, aspect=2.6, text width=4.6em , font=\small, fill=ProcessBlue!15, draw=RoyalBlue},
  term/.style={proc, rounded corners, fill=yellow!60, draw=black!50},
  region/.style={proc, rounded corners, minimum height=11.2cm, minimum width=9cm, fill=gray!10, draw=RoyalPurple},
  region1/.style={proc, rounded corners, minimum height=9.3cm, minimum width=9cm, fill=gray!10, draw=ForestGreen},
  % -------------------------------------------------
  % Connector line styles for different parts of the diagram
  norm/.style={->, draw, black},
  it/.style={font={\small\itshape}}
}

\node[region, font=\rotatebox{90}, text=blue, align=center](r1){};   
\node[rotate=90, font=\Large, yshift = -4mm] at (r1.west) {PPA-focused flow};
\node [longproc, below left =-4.2cm and 1.7cm of r1, it] (p0) {Technology \\ and tool settings};
\node [fproc, join, right=4cm of p0] (p1) {};
\node[] at (p1.center) {floor/power plan};
\node [fproc, below=7mm of p1] (p2) {};
\node[] at (p2.center) {placement};
\node [fproc, below=7mm of p2] (p3) {};
\node[] at (p3.center) {opt (pre-CTS)};
\node [test, below left=7mm and 4cm of p3] (t1) {$slack>0$ ?};
\node [fproc, below=1.7cm of t1] (p4) {};
\node[] at (p4.center) {CTS};
\node [fproc, below=7mm of p4] (p5) {};
\node[] at (p5.center) {opt (post-CTS)};
\node [fproc, below=7mm of p5] (p6) {};
\node[] at (p6.center) {route};
\node [test, above right=1.3cm and 4cm of p6] (t2) {$viol$ ?};
\node [test, below=2cm of t2] (t3) {$slack>0$ ?};
\node [fproc, below left=.7cm and 4cm of t3] (p7) {};
\node[] at (p7.center) {opt (power)};
\node [fproc, below=7mm of p7] (p8) {};
\node[] at (p8.center) {opt (post-route)};
\node [test, join, below=1.7cm of p8] (t4) {$improved$ ?};
\node [term, right=4cm of t4] (final) {routed design};
\node[region1, below right=5.75cm and 1.7cm of t4](r2){};
\node[rotate=90, font=\Large, yshift = -4mm] at (r2.west) {Security-aware flow};
\node [rproc, below=2cm of final] (p9) {};
\node[] at (p9.center) {TI};
\node [test, join, below=2cm of t4] (t5) {$slack>0$ ?};
\node [lproc,  below=2cm of t5] (p10) {opt (post-route)};
\node [test, join, below=2cm of p9] (t55) {$improved$ ?};
\node [rproc,  below=2cm of t55] (p11) {};
\node[] at (p11.center) {FSPFI};
\node [test, join, below=2cm of p10] (t6) {$slack>0$ ?};
\node [lproc, below=2cm of t6] (p12) {opt (post-route)};
\node [test, join, below=2cm of p11] (t7) {$improved$ ?};
\node [term, below left=1.7cm and 1.7cm of t7] (finalized) {finalized design};

\draw [->] (p3.south) -- ++(-3cm,0) |- ++(0,3cm)  -|  (t1.north);

\draw [*->,RoyalBlue] (t1.east) -- ++(5mm,0) -- ++ (0,-2cm) -|  (p3.east);
\draw [*->,RoyalBlue] (t1.south) -- (p4);
\path (t1.east) to node [near start, yshift=-3mm] {$n$} (p3);
\path (t1.south) to node [near start, yshift=-.5em, xshift=2mm] {$y$} (p4);

\draw [*->,RoyalBlue] (t2.west) -- ++(-4mm,0)  -- ++(0,-2cm) |- (p6);
\draw [*->,RoyalBlue] (t2.south) -- (t3);
\path (t2.south) to node [near start, xshift=2mm , yshift=-1mm] {$n$} (t3);
\path (t2.west) to node [near start, xshift=5.5mm, yshift=-1mm] {$y$} (p6);

\draw [->] (p6.east) -- ++(0,-1cm) |- ++(10cm,0)  -| ++(0,15cm) -| ++(10cm,0) -- (t2.north);

\draw [*->,RoyalBlue] (t3.west)  -- ++(-4mm,0)  -- ++(0,-2cm) |- (p7);
\draw [*->,RoyalBlue] (t3.south)  |- (p8);
\path (t3.south) to node [near start, xshift=9mm, yshift=-2mm] {$n$} (p8);
\path (t3.west) to node [near start, , xshift=4mm, yshift=-2.5mm] {$y$} (p7);

\draw [*->,RoyalBlue] (t4.east) -- (final);
\draw [->]  (final.south) -- (p9.south);
\path (t4.west) to node [near start, xshift=-4mm, yshift=0mm] {$y$} (p8);
\path (t4.east) to node [near start, xshift=1mm, yshift=2mm] {$n$} (final);

\draw [*->,RoyalBlue] (t4.west) -- ++(-1.5cm,0) |-  (p8.south);

\draw [*->,RoyalBlue] (t5.south)  -- ++ (0,-18mm) -- ++(30cm,0)  -- ++(0,-2cm) |- (p11.west);
\draw [*->,RoyalBlue] (t5.west) -- ++ (-2cm , 0) |- (p10.west);
\path (t5.west) to node [near start, xshift=-5mm, yshift=1mm] {$n$} (p10);
\path (t5.south) to node [near start, xshift=-6mm, yshift=6mm] {$y$} (p11);

\draw [*->,RoyalBlue] (t55.south)   -- ++ (0, -1.5cm) -| (p10.south);
\draw [*->,RoyalBlue] (t55.east) -- ++ (4mm, 0) -- ++(0, -5.5cm) -| (p11);
\path (t55.south) to node [near start, xshift=3mm, yshift=-2mm] {$n$} (p10);
\path (t55.east) to node [near start, xshift=2mm, yshift=1mm] {$y$} (p11);

 \draw [*->,RoyalBlue] (t6.west)  -- ++ (-2cm , 0) |- (finalized.west);
 \draw [*->,RoyalBlue] (t6.south)  -- (p12);
\path (t6.west) to node [near start, xshift=-8mm, yshift=6mm] {$y$} (finalized.west);
\path (t6.south) to node [near start, xshift=3mm, yshift=-1mm] {$n$} (p12);

 \draw [*->,RoyalBlue] (t7.south)  -- ++(0, -2cm) -| (p12.south);
 \draw [*->,RoyalBlue] (t7.east)  -- ++ (2cm, 0cm ) |- (finalized.east);
 \path (t7.east) to node [near start, xshift=8mm, yshift=0mm] {$n$} (finalized);
\path (t7.south) to node [near start, xshift=4mm, yshift=-3mm] {$y$} (p12);

\end{tikzpicture}
\end{adjustbox}
\caption{Implementation flow utilized in this work along with the security-aware steps.}
\label{fig:flow}
\end{figure}

The flow starts by setting up tool/technology settings and proceeds with the floorplanning and powerplanning. We purposefully do \textbf{not} change tool settings for different designs (as done in prior research). Floorplanning is executed considering the minimum area that still allows for a design to pass timing without DRC violations. Next, placement and pre-clock tree optimizations take place. If setup is violated after optimization, the setup margin is increased by $1ps$ and optimizations start again\footnote{The optimization engine is asked to work harder on every path in $1ps$ increments.}. If timing passes, then the clock tree is built followed by post-CTS optimizations and routing. If there are routing-related DRC violations left, the routing step is repeated\footnote{Calling the routing engine multiple times can solve small routing issues but cannot solve generalized congestion issues.}. Another slack check is performed and if there is a positive timing slack, Innovus is instructed to perform power optimizations. Otherwise, timing-oriented post-route optimizations are performed. Finally, multiple calls to this optimization step are done until no improvement is found and the solution is considered converged and final. 

The output of the IMP strategy becomes the input to the TI strategy that we discuss next. 

\subsection{TI strategy}

Our proposed strategy to counter HTs relies solely on eliminating continuous empty placement sites within a layout. If said sites are not present, the abundance of routing resources to connect an eventual HT becomes irrelevant. The strategy is described in detail in Alg.~\ref{alg:ti} and it is built on the notions of \emph{soft nudges} and \emph{hard pushes} on an existing placement solution. Nudges are the movement of cells located on the periphery of a vulnerable region towards the center of that region. A nudge moves a cell horizontally, keeping it placed in the same row where it was already placed. Nudges are localized and tend to incur little to no impact on PPA and for this reason are preferred over pushes. Pushes are movements in the vertical direction, meaning that cells will be moved up/down one row towards the center of the vulnerable region.

\begin{algorithm}[tb]
\begin{algorithmic}[1]
\Require $layout$ \textbf{Ensure:} $layout$ without vulnerable regions

\State $regions \gets vul\_regions(layout, 20,SITE\_SIZE)$
\State $count \gets size\_of(regions)$, $best\_count \gets \infty$
\While{$count \neq 0$}
    \If{$count < best\_count$} \Comment{Last round was good}
        \State $best\_count \gets count$, $stuck \gets 0$
    \Else \Comment{Last round was bad}
        \State $stuck \gets stuck +1$
    \EndIf

    \State $nudge \gets (stuck < STUCK\_MAX)$ ? $true:false$
    \State $push \gets (stuck < STUCK\_MAX)$ ? $false:true$

    \ForAll{$r \in regions$}
        \State $corners \gets find\_corners(r)$
        \ForAll{$c \in corners$} 
            \State $cell \gets find\_cell\_near(c)$
            \If{$nudge = true$}
                \State $cell \gets nudge(cell, SITE\_SIZE)$
            \EndIf
            \If{$push = true$}
                \State $cell \gets push(cell, SITE\_HEIGHT)$
                \State $layout \gets \textcolor{blue}{eco\_place}(layout)$
            \EndIf
        \EndFor
    \EndFor
    \State $regions \gets vul\_regions(layout, 20, SITE\_SIZE)$
    \State $count \gets size\_of(regions)$
    \EndWhile
\State $layout \gets \textcolor{blue}{opt\_design}(layout)$
    \State return $layout$
\end{algorithmic}
\caption{TI strategy}
\label{alg:ti}
\end{algorithm}

The algorithm starts by finding vulnerable regions within the layout (line 1). According to the ISPD'22 threat model, a vulnerable region must have at least 20 continuous empty placement sites (defined as $SITE\_SIZE$). The number of regions is obtained (line 2) and if that number is greater than zero, the main loop of the algorithm starts (line 3). The variable $best\_count$ (line 2) is used to keep track of the previous lowest count and is updated accordingly (lines 4-5). A variable named $stuck$ (line 5) keeps track of the number of consecutive iterations of the main loop that did not improve the solution (line 7). A constant named $STUCK\_MAX$ (line 8) determines how many consecutive stucks can happen. If the number of stucks is small, the algorithm will perform a nudge (line 8), otherwise, it will perform a push (line 9). The inner loop of the algorithm iterates over all vulnerable regions (line 10), finds their upper right, lower right, upper left, and lower left corners (line 11), and, for each corner, finds the periphery cell closest to it (line 13). That cell will be either nudged by one $SITE\_SIZE$ (line 15) or pushed by one $SITE\_HEIGHT$ (line 17). If a push occurs, Innovus is instructed to execute a round of \textcolor{blue}{eco\_place} to legalize the placement solution. The algorithm finds the remaining vulnerable regions (line 19) and continues with the main loop (line 3). An optimized layout is returned when the loop is aborted (lines 21-22). Keywords in blue correspond to Innovus commands.

\subsection{FSPFI strategy}

Our proposed FSPFI strategy is divided into two phases. In Phase A, we attempt to \emph{push down} any net assets to lower metal layers, giving them a higher chance of being covered by non-asset nets. In Phase B, we apply Non-Default Rules (NDR) to route non-asset nets with \emph{wider metals}, thus increasing the chances that assets are covered by the non-assets. In both phases, we utilize an ECO-styled strategy with progressive rounds while limiting the number of nets that should be rerouted per round. This ECO-like approach helps will convergence and in reducing execution times since it avoids rebuilding the global routing solution.

It should be noted that some designs considered in the ISPD'22 contest have hundreds of net assets, therefore pushing all of them down to lower metal layers is challenging. There are two scenarios that are undesirable: a) pushing a net asset down might force another net asset to be promoted to an upper layer, therefore nullifying the effort; (b) net assets that are pushed down compete for lower metal routing resources with timing-critical nets, which can hurt the performance of the design. For these reasons, Phase A is terminated when the FSPFI exposure is no longer improved. Upon switching to Phase B, the goal is to widen non-net assets as much as possible until routing resources are exhausted and DRCs start to appear. 

Our FSPFI strategy is detailed in Alg.~\ref{alg:fspfi}, which starts by calculating the current exposure (line 1). The main loop of Phase A repeats while exposure is being improved (line 2), and the best exposure value is kept in the $best\_exp$ variable (line 3). Then, for each net asset, its exposure factor is calculated as its area multiplied by its percentage exposure (line 5). The result is stored in the map $factors$ which is then used to rank the net assets (line 6). Net assets are assigned preferred routing layers (lines 8-9). A new layout is generated by rerouting the current solution to respect the new preferred layers (line 10), the respective exposure of the new layout is obtained (line 11). 

\begin{algorithm}[!]
\begin{algorithmic}[1]
%\textbf{Require:} $layout$
\Require $layout$ \textbf{Ensure:} $layout$ with fewer exposed assets

\State $exp \gets find\_exposure(layout), best\_exp \gets \infty$

\While{$exp \leq best\_exp$} \Comment{Phase A}
    \State $best\_exp \gets exp$
    \ForAll{$net \in assets$}
        \State $factors(net) \gets net.area \times net.exp\_perc $
    \EndFor
    \State $assets.sort\_by(factors)$
    \For{$i \gets 1$ to $NETS\_PER\_RD$} 
        \State $assets[i].preferred\_layers.bot \gets M1$
        \State $assets[i].preferred\_layers.top \gets M\_TOP$
    \EndFor
    \State $layout \gets \textcolor{blue}{route\_detail}(layout)$
    \State $exp \gets find\_exposure(layout)$
\EndWhile

\State{$layer \gets TOP\_METAL$}%\Comment{10 for AES, 6 for others}
\While{$\textcolor{blue}{check\_drc}(layout) = pass$} \Comment{Phase B}
    \State{$counter \gets 0$} 
    \ForAll{$net \in nets$}
        \If {$is\_asset(net) = false$}
            \If {$is\_widen(net) = false$}
                \If {$has\_wires\_in(net, layer) = true$}
                    \State $widen(net,layer), incr(counter)$
                    \If{$ct = NETS\_PER\_RD$}
                        \State $\textcolor{blue}{route\_detail}(layout)$
                    \EndIf
                \EndIf
            \EndIf
        \EndIf
    \EndFor
    \If{$counter = 0$}
        \State $layer \gets layer -1$
    \EndIf    
\EndWhile
\State return $layout*$
\end{algorithmic}
\caption{FSPFI strategy}
\label{alg:fspfi}
\end{algorithm}

Phase B of Alg.~\ref{alg:fspfi} starts by setting the layer of interest $layer$ for widening (line 12) and it continues while DRC violations are not created (line 13). The $counter$ variable is used to limit the number of nets to be rerouted per round (line 14). For each non-asset net (line 16) that has not been widened before (line 17) and that has wires in the $layer$ (line 18), widening is performed on the net $net$ and the counter is incremented (line 19). The \textcolor{blue}{route\_detail} command is called whenever the counter reaches its upper limit (lines 20-21). If not enough nets exist, the $layer$ is decreased by 1 (line 23). Finally, the algorithm returns $layout*$, which is the layout obtained in the last round that did not have DRC violations.

\pgfplotstableread[row sep=\\,col sep=&,columns/Benchmark/.style={string type}]{
Benchmark& imp & tistrat & fspfistrat  \\
    AES\_1        & 0.42776 & 0.42777 & 0.42847 \\
    AES\_2        & 0.39869 & 0.39872 & 0.39959 \\
    AES\_3        & 0.45317 & 0.45314 & 0.45608 \\
    CAMELLIA      & 0.37959 & 0.37902 & 0.37919 \\
    CAST          & 0.37978 & 0.37944 & 0.38006 \\
    MISTY         & 0.36691 & 0.36684 & 0.36710 \\
    OMSP430\_1    & 0.39305 & 0.39308 & 0.39365 \\
    OMSP430\_2    & 0.44688 & 0.44699 & 0.44735 \\
    PRESENT       & 0.33348 & 0.33365 & 0.33390 \\
    SEED          & 0.37733 & 0.37730 & 0.37466 \\
    SPARX         & 0.37467 & 0.37477 & 0.37660 \\
    TDEA          & 0.44294 & 0.44276 & 0.44445 \\
}\desmetric 

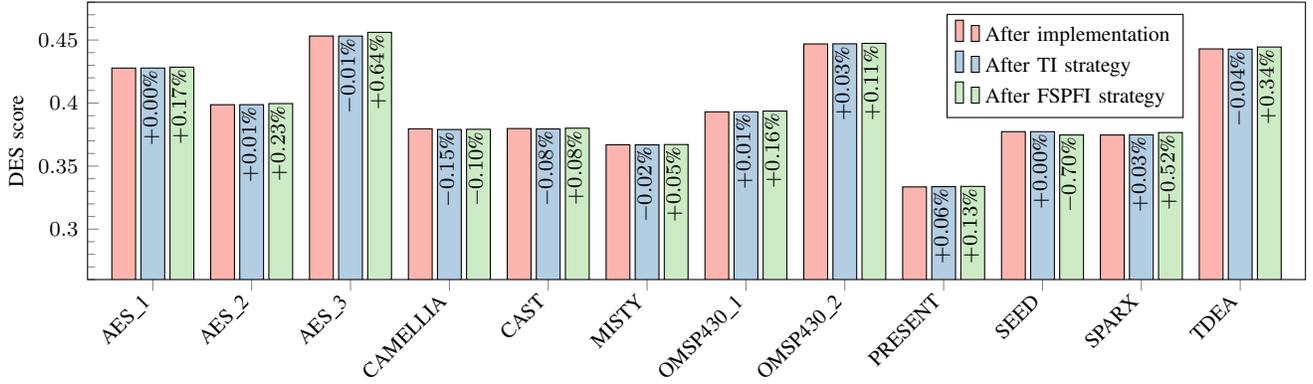
\begin{figure*}[!]
\centering
\begin{tikzpicture}
    \begin{axis}[
        point meta = explicit symbolic,
        ybar,
        enlarge x limits=0.06,
        symbolic x coords={AES\_1, AES\_2, AES\_3, CAMELLIA, CAST, MISTY, OMSP430\_1, OMSP430\_2, PRESENT, SEED, SPARX, TDEA},
        x tick label style={rotate=45, anchor=east,font=\footnotesize},
        y tick label style={font=\footnotesize},
        xtick pos=left,
        ytick pos=left,
        ylabel style={font=\footnotesize, yshift=-8pt},
        width = .98\linewidth,
        height = 150pt,
        minor y tick num = 4,
        ymin = 0.26,
        ymax = 0.48,
        ylabel=DES score,
        legend style={at={(0.9,0.77)},anchor=east,font=\footnotesize},
        legend cell align=left,        
        visualization depends on={\thisrow{imp} \as \myimp},
        visualization depends on={\thisrow{tistrat} \as \myti},
        visualization depends on={\thisrow{fspfistrat} \as \myfspfi},
        bar width=9pt,
        ]
   
        \addplot [fill=mycolor1] table[x=Benchmark,y=imp]{\desmetric};
        \addplot[fill=mycolor2, nodes near coords, nodes near coords={\pgfkeys{/pgf/number format/.cd,fixed, showpos, fixed zerofill,precision=2}\pgfmathparse{(\myti*100 / \myimp)-100}\pgfmathprintnumber\pgfmathresult\%}, nodes near coords style={font=\footnotesize, rotate=90, xshift=-15pt, yshift=-6pt}] table[x=Benchmark,y=tistrat]{\desmetric};
        \addplot[fill=mycolor3, nodes near coords, nodes near coords={\pgfkeys{/pgf/number format/.cd,fixed, showpos, fixed zerofill,precision=2}\pgfmathparse{(\myfspfi*100 / \myimp)-100}\pgfmathprintnumber\pgfmathresult\%}, nodes near coords style={font=\footnotesize, rotate=90, xshift=-15pt, yshift=-6pt}] table[x=Benchmark,y=fspfistrat]{\desmetric};
       
        \addlegendentry{After implementation}
        \addlegendentry{After TI strategy}
        \addlegendentry{After FSPFI strategy}
    \end{axis}
\end{tikzpicture}
% \end{adjustbox}
\caption{DES scores as the design evolves.}
\label{fig:des}
\end{figure*}

\section{Results} \label{sec:results}

All experiments reported in this section are executed in Cadence Innovus v21.16 utilizing the Nangate 45nm Open Cell library~\cite{nangate45}. The experiments were executed on a server equipped with an Intel(R) Xeon(R) Silver 4208 CPU and 128GB of memory. The chosen circuits are the same twelve benchmarks from the ISPD'22 competition as well as an additional \emph{large} design that contains 64 instances of the same processor that operate in lockstep. We emphasize again that our implementation flow is design-agnostic and that no tool settings are changed to obtain improved scores for one specific benchmark. The values adopted for the many constants are given in Tab.~\ref{tab:constants}. The values for $STUCK\_MAX$ and $NETS\_PER\_RD$ determine how quickly the TI and FSPFI strategies converge, respectively. In general, too small values increase runtime while too large values lead to non-convergence. A sensitivity analysis for these constants is reported in Section~\ref{sec:sensitivity}. The values for $SITE\_SIZE$ and $SITE\_HEIGHT$ are a property of the technology and should be set accordingly. Finally, the values for $M\_TOP$ and $TOP\_METAL$ are a property of the metal stack and are set according to the contest settings: AES designs use a 10-metal stack, and all others use a 6-metal stack.

\begin{table}[htbp]
    \caption{Constants utilized in the TI/FSPFI algorithms.}
    \label{tab:constants}
    \renewcommand{\arraystretch}{1.3}
    \begin{threeparttable}[ht] {
        \setlength{\tabcolsep}{15pt}
        \renewcommand{\TPTminimum}{\linewidth}
        \makebox[\linewidth]{%
            \begin{tabular}{lc}
                \hline
                \textbf{Constant} & \textbf{Value} \\
                \hline
                \rowcolor[HTML]{EFEFEF} $SITE\_SIZE$ &  0.19 \si{\micro\meter} \\
                $SITE\_HEIGHT$ & 1.4 \si{\micro\meter} \\
                \rowcolor[HTML]{EFEFEF} $STUCK\_MAX$ & 3  \\
                $NETS\_PER\_RD$ & 0.05\% of total \\
                \rowcolor[HTML]{EFEFEF} $M\_TOP$ & 4 (AES) - 3 (others)\\
                % $M\_TOP$ & 3 (others)  \\ 
                 $TOP\_METAL$ & 10 (AES) - 6 (others) \\
                % $TOP\_METAL$ & 6 (others)\\
\hline
            \end{tabular}
        }
    }
    \end{threeparttable}
\end{table}

All results presented in this section are \textbf{DRC-clean} and \textbf{pass timing}. For this reason, we do not provide DRC count (always zero) or timing slack (always positive) for any of the considered benchmarks.

Our main claim of zero-overhead security closure is visually confirmed from Fig.~\ref{fig:des} where we track the Design Quality (DES as defined in Eq.~\ref{eq1}) as the design goes through our three strategies. The percentage values annotated on the colored bars represent the overhead with respect to the IMP strategy that is not security-aware. The worst-case overhead is a 0.64\% increase for AES\_3 after the FSPFI strategy is applied, but this number is a clear outlier. The average overhead of the TI strategy is $-0.13\%$. The average overhead of the FSPFI strategy is $+0.13\%$. Note that there are several cases of negative overheads, a clear indication that the values we report are within the margins of the heuristic behavior of the P\&R engine.

A detailed overview of several metrics related to the applied strategies is given in Tab.~\ref{tab:completeresults}. First, note that density remains nearly the same after TI and FSPFI strategies are applied, indicating that our security-related strategies do not compromise on area. Also, note that all vulnerability regions are solved completely after the TI strategy is applied. Also important to highlight is that the TI strategy does not affect the FSPFI score, indicating that there is no convergence issue between TI  and FSPFI strategies: it is always possible to run TI before FSPFI with no overhead to PPA or security. The last column of Tab.~\ref{tab:completeresults} is the overall score obtained by applying Eq.~\ref{eq1}.

\begin{table*}[tbp]
\caption{Final results after implementation, TI, and FSPFI strategies.}
\label{tab:completeresults}
\begin{threeparttable}[ht]
\renewcommand{\TPTminimum}{\linewidth}
\renewcommand{\arraystretch}{1.3}
\makebox[\linewidth]{%
\label{tab:final}
\begin{tabular}{l|lll|llll|lllll}
\hline
\multicolumn{1}{c|}{\textbf{Benchmark}} & \multicolumn{3}{c|}{\textbf{After implementation}} &  \multicolumn{4}{c|}{\textbf{After TI strategy}}  & \multicolumn{5}{c}{\textbf{After FSPFI strategy}}     \\ 
 & \textbf{Density} & \textbf{VRs} & \textbf{FSPFI} & \textbf{N/P (C)} &\textbf{Density} & \textbf{VRs} & \textbf{FSPFI} & \textbf{Density} & \textbf{R (A+B)} & \textbf{ECO'd (A+B)} & \textbf{FSPFI} & \textbf{Overall} \\ \hline
\rowcolor[HTML]{EFEFEF} AES\_1 & 95.68\% & 3 & 0.78512 & 2/0 (7)   & 95.67\% & 0 & 0.78527 & 95.72\% & 2+7 & 119+325 & 0.59026 & 0.14563 \\
                        AES\_2 & 95.92\% & 4 & 1.13439 & 2/0 (10)   & 95.93\% & 0 & 1.13446 & 96.02\% & 2+14 & 120+491 & 0.95279 & 0.19036  \\
\rowcolor[HTML]{EFEFEF} AES\_3 & 95.61\% & 4 & 0.84763 & 4/0 (13)   & 95.60\% & 0 & 0.84749 & 95.71\% & 5+16 & 123+636 & 0.68821 & 0.15694  \\
                        CAMELLIA & 96.30\% & 3 & 0.80760 & 13/0 (6)  & 96.09\% & 0 & 0.80744 & 96.10\% & 3+3 & 50+108 & 0.77673 & 0.14726  \\
\rowcolor[HTML]{EFEFEF} CAST & 94.21\% & 1 & 0.80561 & 1/0 (1) & 94.15\% & 0 & 0.80547 & 94.15\% & 4+3 & 94+195 & 0.79969 & 0.15196 \\
                        MISTY & 95.59\% & 5 & 0.76450 & 14/1 (24) & 95.55\% & 0 & 0.76451 & 95.60\% & 1+4 & 14+196 & 0.75313 & 0.13824 \\ 
\rowcolor[HTML]{EFEFEF} OMSP430\_1 & 95.95\% & 5 & 0.86478 & 4/0 (9) & 95.95\% & 0 & 0.86500 & 95.95\% & 7+78 & 23+277 & 0.83653 & 0.16465 \\
                        OMSP430\_2 & 94.18\% & 15 & 0.97727 & 23/1 (52) & 94.18\% & 0 & 0.97855 & 94.18\% & 7+67 & 18+258 & 0.94492 & 0.21135 \\ 
\rowcolor[HTML]{EFEFEF} PRESENT & 97.35\% & 1 & 0.84591 & 15/1 (14)   & 97.31\% & 0 & 0.84753 & 97.31\% & 5+27 & 12+84 & 0.76771 & 0.12817 \\
                        SEED & 94.14\% & 5 & 0.78873 & 20/2 (21) & 94.10\% & 0 & 0.78933 & 93.66\% & 5+4 & 129+256 & 0.76308 & 0.14295 \\
\rowcolor[HTML]{EFEFEF} SPARX & 98.06\% & 5 & 0.88972 & 38/3 (58) & 98.05\% & 0 & 0.89190 & 98.09\% & 3+13 & 74+545 & 0.71919 & 0.13542 \\ 
                        TDEA & 94.57\% & 2 & 0.92258 & 6/0 (6)   & 94.47\% & 0 & 0.92324 & 94.55\% & 7+59 & 29+511 & 0.75098 & 0.16689  \\ \hline
                                 % sparx is the highest density increase. likely related to vertical pushes?
\end{tabular}
}
 \begin{tablenotes}
      \item \footnotesize \textbf{VRs} = vulnerable regions. \textbf{N/P} = nudges/pushes. \textbf{(C)} total number of cells touched during TI strategy. \textbf{R} = rounds. \textbf{A+B} = phases of the FSPFI strategy. 
\end{tablenotes}
\end{threeparttable}
\end{table*}

\begin{figure*}[h]
    \centering
    \begin{adjustbox}{width=1\textwidth}

    \subfloat[]{
        \begin{tikzpicture}[spy using outlines={circle,black,magnification=4,size=1.4cm, connect spies}]
            \node {\includegraphics[width=0.45\columnwidth, trim=0cm 0cm 0cm 0cm]{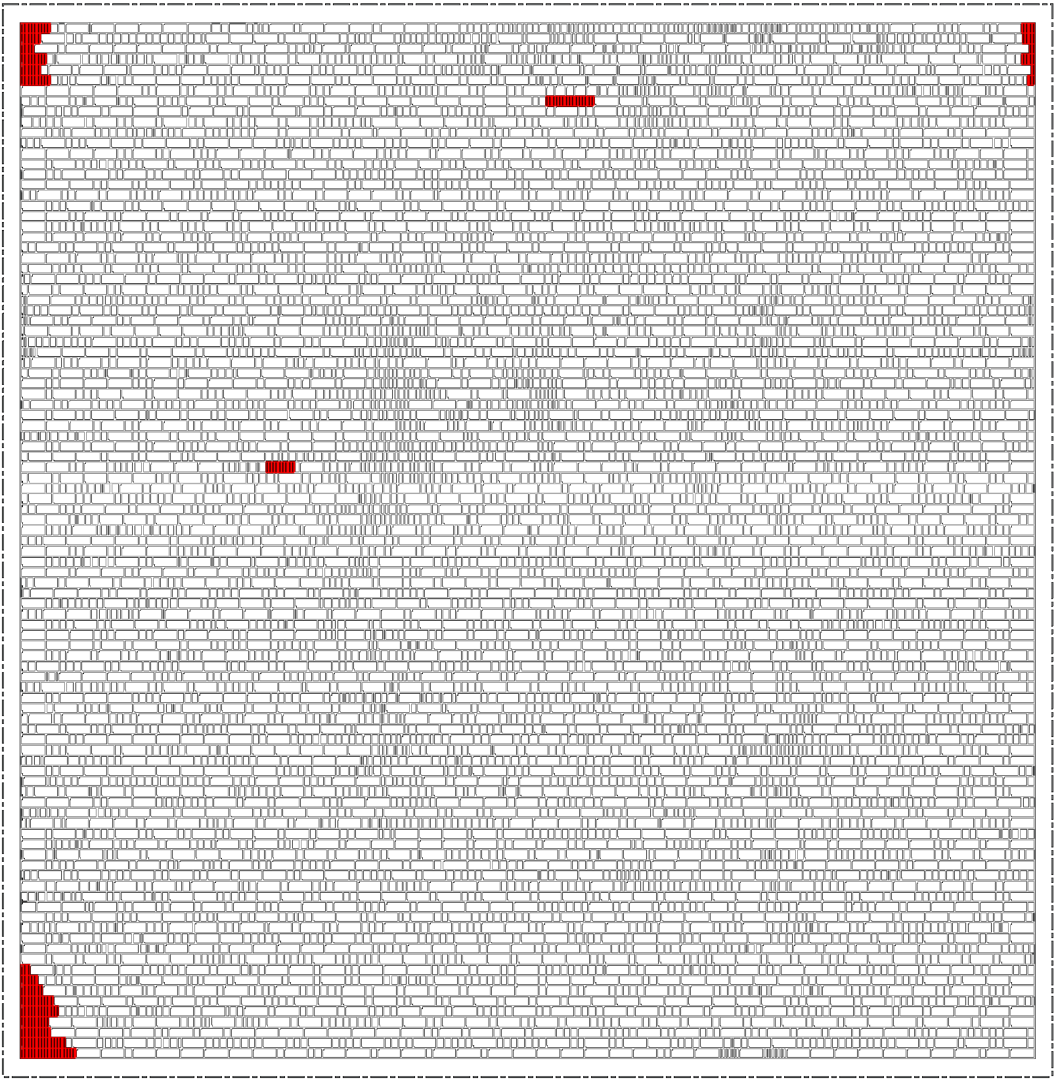}};
            \spy on (-1.80,-1.83) in node [right] at (-3.5,0);
        \end{tikzpicture}
    }
    \subfloat[]{
        \begin{tikzpicture}[spy using outlines={circle,black,magnification=4,size=1.4cm, connect spies}]
            \node {\includegraphics[width=0.45\columnwidth, trim=0cm 0cm 0cm 0cm]{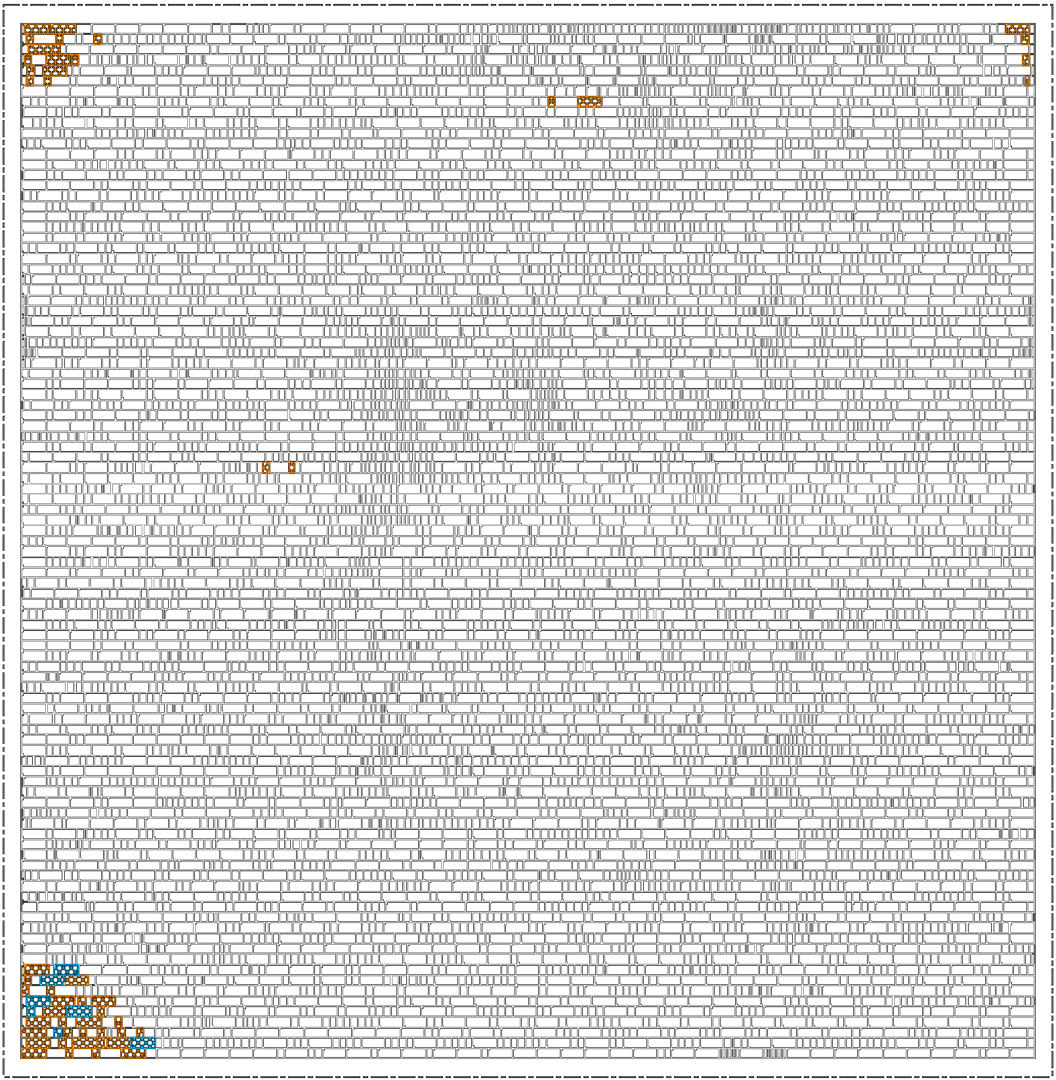}};
            \spy on (-1.80,-1.83) in node [right] at (-3.5,0);
        \end{tikzpicture}
    }    
    \hspace{30pt}
    \subfloat{%
       \raisebox{60pt}[0pt][0pt]{      
        \parbox{0.35\linewidth}{
        \begin{shaded}
            \footnotesize
            \texttt{Red shapes = \textbf{379} sites \\
            Vulnerable areas = \textbf{5} \\
            Largest vulnerable area = \textbf{192} sites \\
            Average vulnerable area = \textbf{75.80} sites \\
            Nudge rounds = \textbf{38} \\
            Push rounds = \textbf{3} \\            
            Brown cells = \textbf{51} \\
            Blue cells = \textbf{7} \\
            Total cells nudged/pushed = \textbf{58}}
        \end{shaded}
        }
        }
    }
    \end{adjustbox}
    \caption{Layouts for SPARX, measuring 140.4 \si{\micro\meter} by 143.4 \si{\micro\meter}. Wires are omitted for clarity. Red areas are empty regions that are vulnerable to the TI threat. Brown cells are nudged horizontally. Blue cells are pushed vertically. (a) Layout before TI strategy; (b) Layout after TI strategy, containing zero TI vulnerable zones}.
    \label{fig:tistrat}
\end{figure*}

A visual representation of the TI strategy is given in Fig.~\ref{fig:tistrat} for the SPARX benchmark. This design was picked for further analysis because it required the highest number of nudges and pushes among all studied benchmark circuits. The layout contains 5 regions deemed vulnerable. As it typically is the case, vulnerable regions are often located on the corners of the layout. In order to completely solve the TI-related vulnerable regions for SPARX, 38 horizontal nudges and 3 vertical pushes were required. 

\colorlet{shadecolor}{gray!10}

\begin{figure*}[htbp]
    \centering
    \begin{adjustbox}{width=1\textwidth}

    \subfloat[]{    
    \begin{tikzpicture}[spy using outlines={circle,black,magnification=4,size=1.4cm, connect spies}]
            \node {\includegraphics[width=0.45\columnwidth, trim=0cm 0cm 0cm 0cm]{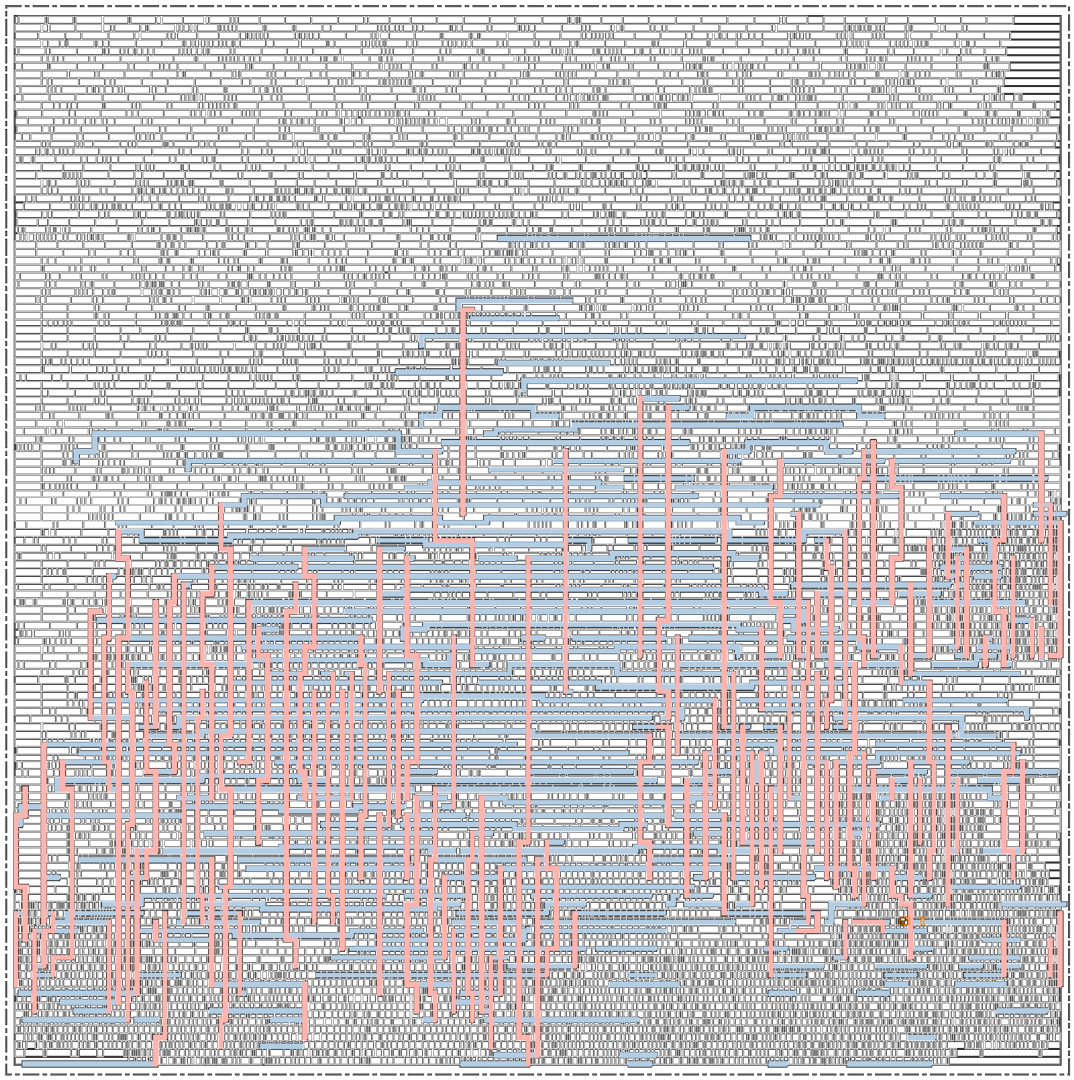}};
            \spy on (0,-0.5) in node [right] at (-2.2,1.4);
        \end{tikzpicture}
        \label{fig:before_reroute}
    }%\hfill
    \subfloat[]{    
    \begin{tikzpicture}[spy using outlines={circle,black,magnification=4,size=1.4cm, connect spies}]
            \node {\includegraphics[width=0.45\columnwidth, trim=0cm 0cm 0cm 0cm]{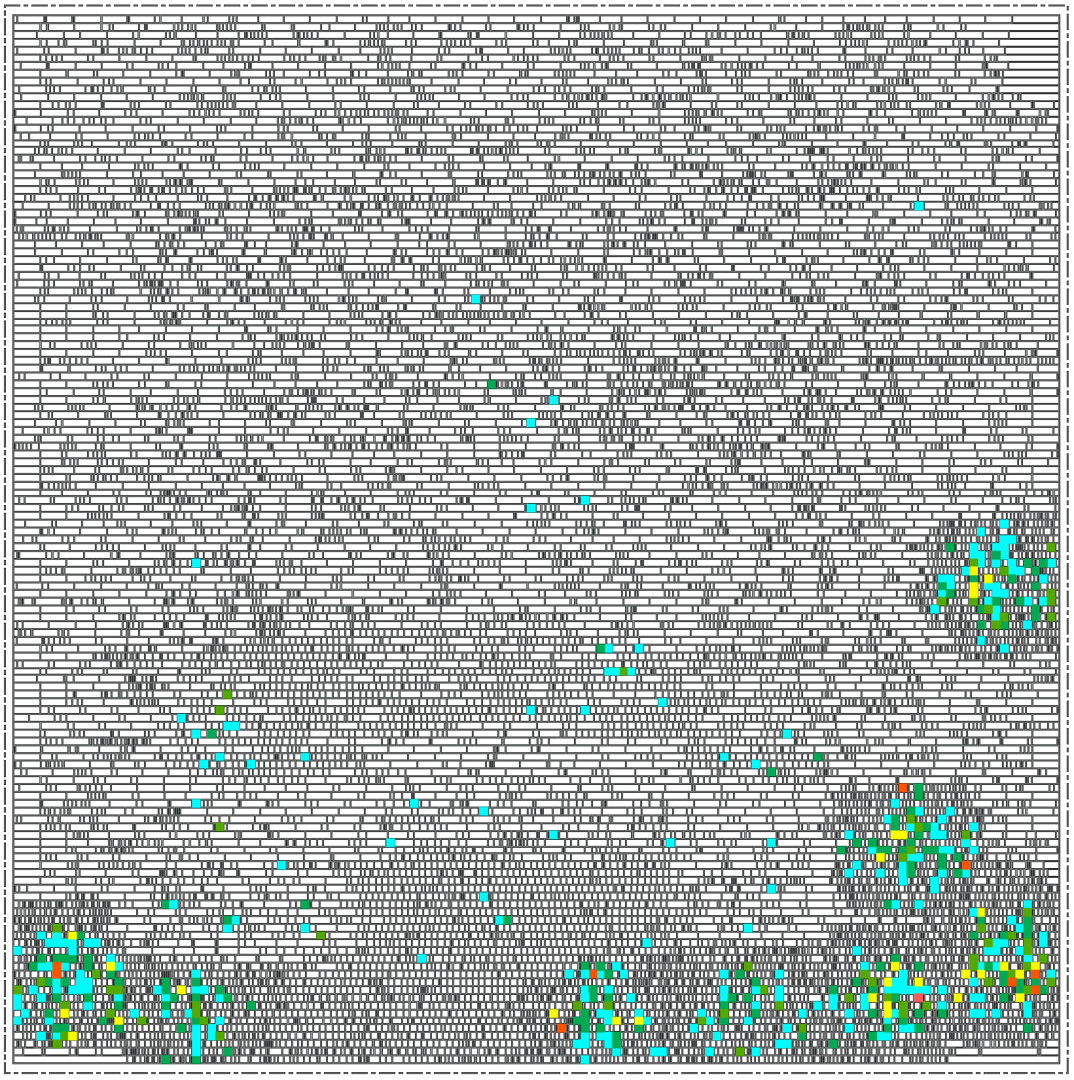}};
        \end{tikzpicture}
        \label{fig:before_reroute_density}
    }%\hfill
    \subfloat[]{    
    \begin{tikzpicture}[spy using outlines={circle,black,magnification=4,size=1.4cm, connect spies}]
            \node {\includegraphics[width=0.45\columnwidth, trim=0cm 0cm 0cm 0cm]{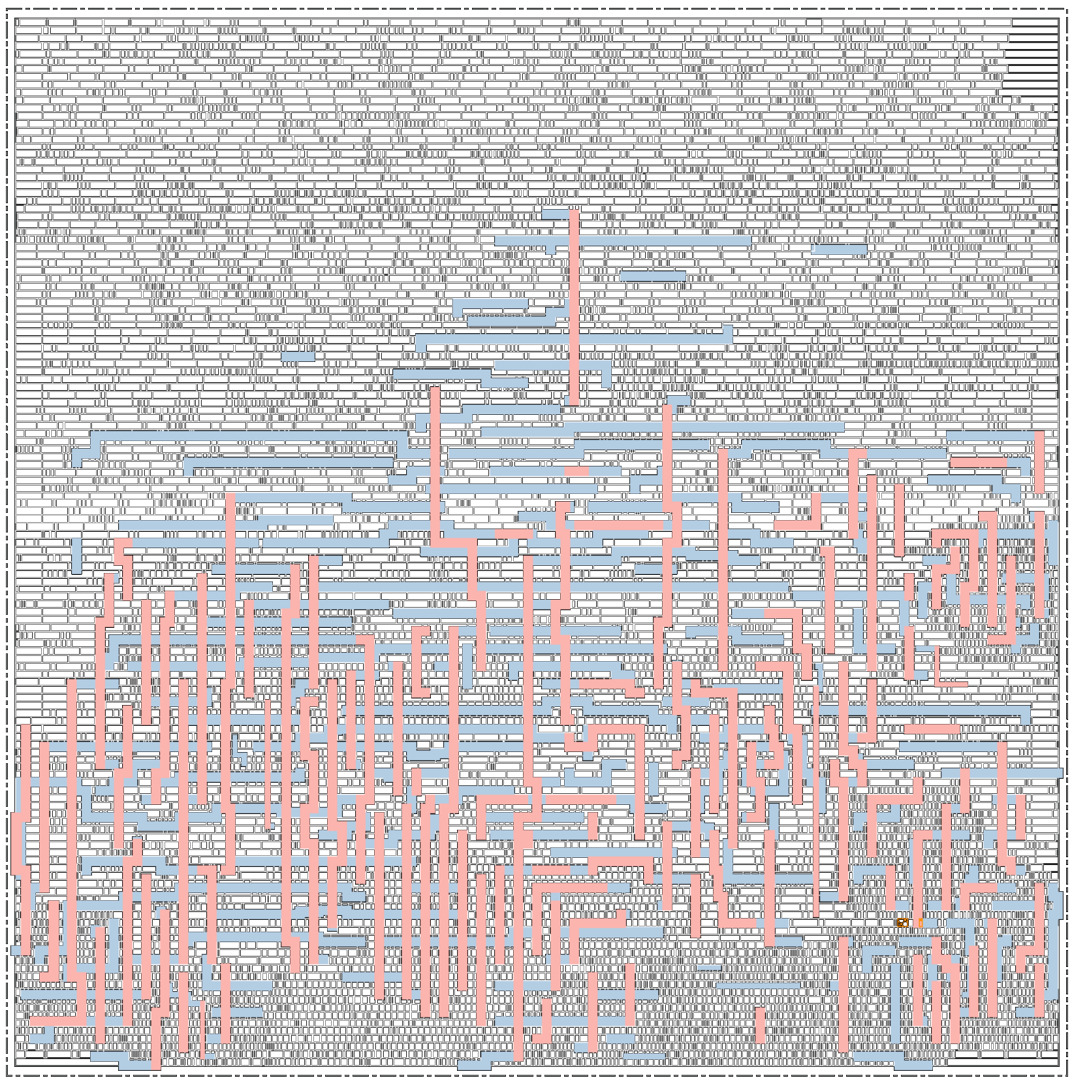}};
            \spy on (0,-0.5) in node [right] at (-2.2,1.4);
        \end{tikzpicture}
        \label{fig:after_reroute}
    }%
    \subfloat[]{    
    \begin{tikzpicture}[spy using outlines={circle,black,magnification=4,size=1.4cm, connect spies}]
            \node {\includegraphics[width=0.45\columnwidth, trim=0cm 0cm 0cm 0cm]{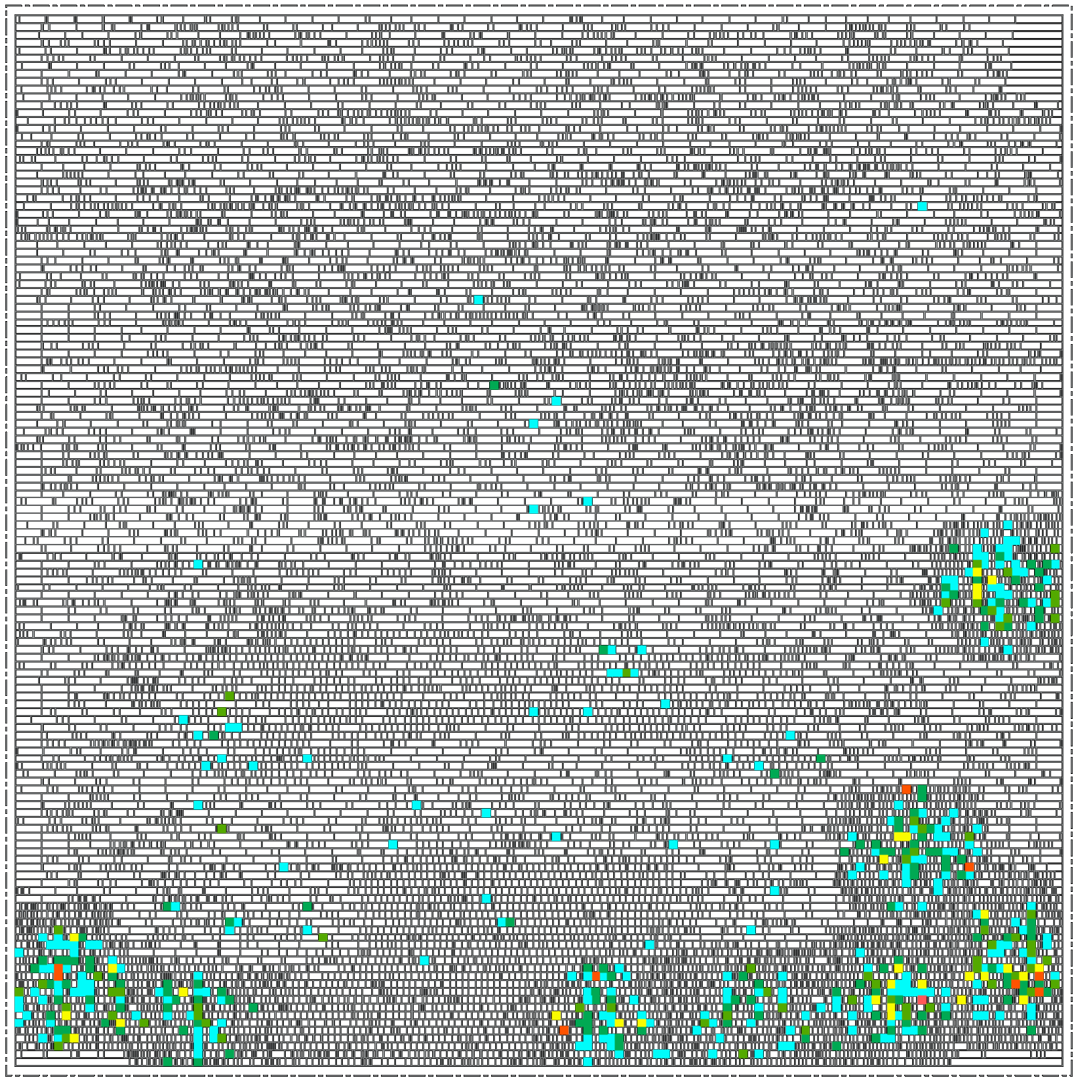}};
        \end{tikzpicture}
        \label{fig:after_reroute_density}
    }%
        \end{adjustbox}
    \caption{Layouts for AES\_2, measuring 191.6 \si{\micro\meter} by 192.4 \si{\micro\meter}. (a) Layout before FSPFI strategy; (c) Layout after FSPFI strategy. Except for \textbf{\textcolor{mycolor1}{M10}} and \textbf{\textcolor{mycolor2}{M9}}, all other layers are omitted. (b) Routing congestion before FSPFI strategy; (d) Routing congestion after FSPFI strategy. Rectangles correspond to grid cells measuring 1.4\si{\micro\meter} by 1.4\si{\micro\meter} and are colored according to routing resource availability: \textcolor{red}{-5}, \textcolor{orange}{-4}, \textcolor{yellow}{-3}, \textcolor{LimeGreen}{-2}, \textcolor{ForestGreen}{-1}, \textcolor{cyan}{-0}.}
    \label{fig:fspfistrat}
\end{figure*}

A visual representation of the FSPFI strategy is given in Fig.~\ref{fig:fspfistrat} for the AES\_2 benchmark in panels (a) and (c). Note how the light blue and pink wires are widened and that more net assets can be hidden under them. Also note that, even after 491 nets were widened, the design remains routable and not overly congested. Panels (b) and (d) show the congestion analysis before and after the FSPFI strategy is applied. It is noteworthy that no new hotspots for routing are created since the wires that are manipulated are in the upper layers, whereas the existing congestion is related to pin access on layers 2 and 3 -- this is validated by noting that the congestion zones correspond to areas with high-density clusters of standard cells. This behavior is representative of other designs that were considered.

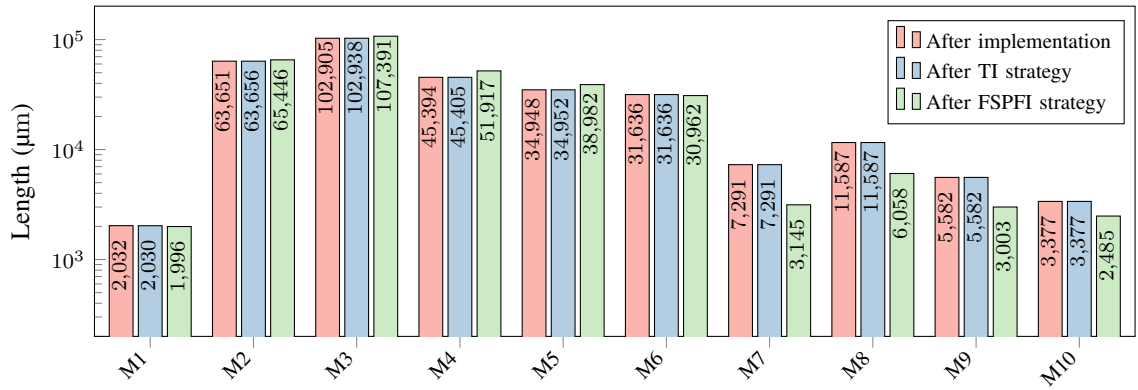
\begin{figure*}[htbp]
\centering
%\begin{adjustbox}{width=0.91\textwidth}
\begin{tikzpicture}
    \begin{axis}[
        point meta = explicit symbolic,
        ybar,
        ymode=log,
        enlarge x limits=0.06,
        symbolic x coords={M1, M2, M3, M4, M5, M6, M7, M8, M9, M10},
        x tick label style={rotate=45, anchor=east,font=\footnotesize},
        y tick label style={font=\footnotesize},
        xtick pos=left,
        ytick pos=left,
        ylabel style={ yshift=-8pt},
        width = .85\linewidth,
        height = 150pt,
        minor y tick num = 4,
        ymin = 200,
        ylabel=Length (\si{\micro\meter}),
        legend style={at={(0.99,0.8)},anchor=east,font=\footnotesize},
        legend cell align=left,        
        visualization depends on={\thisrow{imp} \as \myyimp},
        visualization depends on={\thisrow{tistrat} \as \myyti},
        visualization depends on={\thisrow{fspfistrat} \as \myyfspfi},
        bar width=9pt,
        ]

        \addplot[fill=mycolor1, nodes near coords, nodes near coords={\pgfkeys{/pgf/number format/.cd,fixed, fixed zerofill, precision=0}\pgfmathparse{(\myyimp)}\pgfmathprintnumber\pgfmathresult}, nodes near coords style={font=\footnotesize, rotate=90, xshift=-15pt, yshift=-7pt}] table[x=layer,y=imp]{\wirelength};
        \addplot[fill=mycolor2, nodes near coords, nodes near coords={\pgfkeys{/pgf/number format/.cd,fixed, fixed zerofill, precision=0}\pgfmathparse{(\myyti)}\pgfmathprintnumber\pgfmathresult}, nodes near coords style={font=\footnotesize, rotate=90, xshift=-15pt, yshift=-7pt}] table[x=layer,y=tistrat]{\wirelength};
        \addplot[fill=mycolor3, nodes near coords, nodes near coords={\pgfkeys{/pgf/number format/.cd,fixed, fixed zerofill, precision=0}\pgfmathparse{\myyfspfi)}\pgfmathprintnumber\pgfmathresult}, nodes near coords style={font=\footnotesize, rotate=90, xshift=-15pt, yshift=-7pt}] table[x=layer,y=fspfistrat]{\wirelength};

        %\addplot [fill=mycolor2]  table[x=layer,y=tistrat]{\wirelength};
        %\addplot [fill=mycolor3] table[x=layer,y=fspfistrat]{\wirelength};
             
        \addlegendentry{After implementation}
        \addlegendentry{After TI strategy}
        \addlegendentry{After FSPFI strategy}
    \end{axis}
\end{tikzpicture}
% \end{adjustbox}
\caption{Wirelength per metal layer for the AES\_2 benchmark.}
\label{fig:wirelength}
\end{figure*}

Furthermore, we have collected wirelength statistics for the same AES\_2 benchmark, which are presented in Fig.~\ref{fig:wirelength}. Notice that, as expected, the TI strategy barely changes the metal usage across layers. This is due to two reasons: first, the TI strategy focuses on placing cells in different locations but within the same neighborhood, therefore any affected net is extended by very little length, if any at all. Second, the TI strategy is localized like ECO is, therefore the majority of nets in the design is never rerouted. %could add total number of nets here to make the point clearer

Another trend that is possible to visualize in Fig.~\ref{fig:wirelength} is that the wires are redistributed across metal layers. This is explained by two different effects that take place at the same time: net assets are pushed to lower layers, creating a small increase in wirelength in middle layers like M3, M4, and M5. In turn, regular nets that are not an asset are widened in higher metal layers like M8, M9, and M10. Widened nets take more routing tracks, therefore there are less resources for routing in those layers. In tandem, these two effects cause marginal bloating in middle layers and ease congestion on higher layers. The highest increase seen is 14.3\% in M4, while the highest decrease seen is 56.9\% in M7.

\pgfplotstableread[row sep=\\,col sep=&,columns/Benchmark/.style={string type}]{
layer & imp & tistrat & fspfistrat  \\
    M1 & 2032 & 2030 & 1996 \\
    M2 & 63651 & 63656 & 65446 \\
    M3 & 102905 & 102938 & 107391 \\
    M4 & 45394 & 45405 & 51917 \\
    M5 & 34948 & 34952 & 38982 \\
    M6 & 31636 & 31636 & 30962 \\
    M7 & 7291 & 7291 & 3145 \\
    M8 & 11587 & 11587 & 6058 \\
    M9 & 5582 & 5582 & 3003 \\
    M10 & 3377 & 3377 & 2485 \\
}\wirelength

\pgfplotstableread[row sep=\\,col sep=&,columns/Benchmark/.style={string type}]{
layer & imp & tistrat & fspfistrat  \\
    V1 & 63351 & 63350 & 63546 \\
    V2 & 55287 & 55303 & 57447 \\
    V3 & 20547 & 20556 & 23632 \\
    V4 & 8054 & 8054 & 9929 \\
    V5 & 5463 & 5463 & 6894 \\
    V6 & 1381 & 1381 & 1266 \\
    V7 & 1153 & 1153 & 941 \\
    V8 & 497 & 497 & 433 \\
    V9 & 236 & 236 & 251 \\
}\viacount

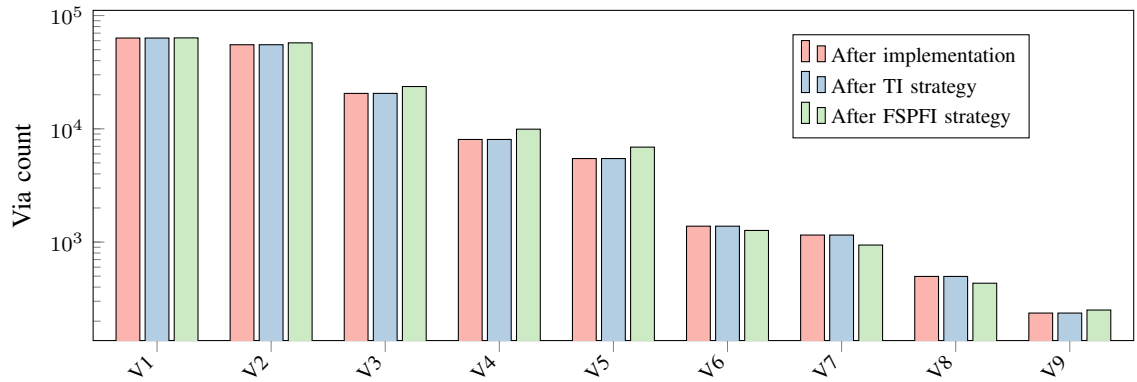
\begin{figure*}[!]
\centering
%\begin{adjustbox}{width=0.91\textwidth}
\begin{tikzpicture}
    \begin{axis}[
        point meta = explicit symbolic,
        ybar,
        ymode=log,
        enlarge x limits=0.07,
        symbolic x coords={V1, V2, V3, V4, V5, V6, V7, V8, V9},
        x tick label style={rotate=45, anchor=east,font=\footnotesize},
        y tick label style={font=\footnotesize},
        ytick pos=left,
        xtick pos=left,
        ylabel style={yshift=-8pt},
        width = .85\linewidth,
        height = 150pt,
        minor y tick num = 4,
        ylabel=Via count,
        legend style={at={(0.9,0.77)},anchor=east,font=\footnotesize},
        legend cell align=left,        
        visualization depends on={\thisrow{imp} \as \myimp},
        visualization depends on={\thisrow{tistrat} \as \myti},
        visualization depends on={\thisrow{fspfistrat} \as \myfspfi},
        bar width=9pt,
        ]
        \addplot [fill=mycolor1]  table[x=layer,y=imp]{\viacount};
        \addplot [fill=mycolor2]  table[x=layer,y=tistrat]{\viacount};
        \addplot [fill=mycolor3] table[x=layer,y=fspfistrat]{\viacount};

        \addlegendentry{After implementation}
        \addlegendentry{After TI strategy}
        \addlegendentry{After FSPFI strategy}
    \end{axis}
\end{tikzpicture}
% \end{adjustbox}
\caption{Via count per layer for the AES\_2 benchmark. VX refers to the via between metal X+1 and metal X.}
\label{fig:viacount}
\end{figure*}

Figure~\ref{fig:viacount} depicts the via count for different layers as the AES\_2 design goes through the different strategies we propose. Note that the number of vias remains the same from implementation to TI strategy, which is expected since no new nets are created. As was the case with the wirelength analysis shown earlier, a redistribution can be seen towards the middle layers. The highest increase in vias happens between M6 and M5, with the number of vias increasing from 5463 to 6894, corresponding to an increase of 26.1\%.

\begin{figure*}[htbp]
\centering
% \begin{adjustbox}{width=0.9\textwidth}
\subfloat[Vulnerable sites for AES\_2]{
\begin{tikzpicture}
    \begin{axis}[
        xlabel near ticks,
        ylabel near ticks,
        xlabel = Round,
        ylabel=Vulnerable sites,
        x tick label style={font=\footnotesize},
        y tick label style={rotate=45, anchor=east, font=\footnotesize},
        ylabel style={font=\footnotesize},
        xlabel style={font=\footnotesize},
        width =  0.33\linewidth,
        ymin = 0,     
        ]   
        \addplot [scatter,
        scatter/classes={nudge={blue, mark=star}, push={red, mark=star},none={mark=star}},
        scatter src=explicit symbolic,
        ] table [x=round, y=sites, meta=style]{\vulnerableaestwo}; 
        \end{axis}
\end{tikzpicture}
}%
\subfloat[Vulnerable sites for SPARX]{
\begin{tikzpicture}
    \begin{axis}[
        xlabel near ticks,
        ylabel near ticks,
        xlabel = Round,
        %ylabel=Vulnerable sites,
        x tick label style={font=\footnotesize},
        y tick label style={rotate=45, anchor=east, font=\footnotesize},
        ylabel style={font=\footnotesize},
        xlabel style={font=\footnotesize},
         width =  0.33\linewidth,
        ymin = 0,      
        legend pos=north east,
        legend style={font=\footnotesize},
        legend cell align=left,
        ]   
        \addlegendimage{only marks, mark=star,blue}
        \addlegendentry{nudge}
        \addlegendimage{only marks, mark=star,red}
        \addlegendentry{push}

        \addplot [scatter, empty legend,
        scatter/classes={nudge={blue, mark=star}, push={red, mark=star},none={mark=star}},
        scatter src=explicit symbolic,
        ] table [x=round, y=sites, meta=style]{\vulnerablesparx};
    \end{axis}
\end{tikzpicture}
}%
\subfloat[Vulnerable sites for OMSP430\_ARRAY]{
\begin{tikzpicture}
    \begin{axis}[
        xlabel near ticks,
        ylabel near ticks,
        xlabel = Round,
        %ylabel=Vulnerable sites,
        x tick label style={font=\footnotesize},
        y tick label style={font=\footnotesize},
        y tick label style={rotate=45, anchor=east, font=\footnotesize},
        xlabel style={font=\footnotesize},
         width =  0.33\linewidth,
        ymin = 0,      
        legend pos=north east,
        legend style={font=\footnotesize},
        legend cell align=left,
        ]   
        \addlegendimage{only marks, mark=star,blue}
        \addlegendentry{nudge}
        \addlegendimage{only marks, mark=star,red}
        \addlegendentry{push}

        \addplot [scatter, empty legend,
        scatter/classes={nudge={blue, mark=star}, push={red, mark=star},none={mark=star}},
        scatter src=explicit symbolic,
        ] table [x=round, y=sites, meta=style]{\vulnerablearray};
    \end{axis}
\end{tikzpicture}
}

\subfloat[Exposure for AES\_2]{
\begin{tikzpicture}
    \begin{axis}[
        smooth,
        axis y line*=left,
        xlabel near ticks,
        ylabel near ticks,
        xlabel = Round,
        x tick label style={font=\footnotesize},
        y tick label style={rotate=45, anchor=east, font=\footnotesize},
        ylabel style={font=\footnotesize},
        xlabel style={font=\footnotesize},
        width =  0.31\linewidth,
        ylabel=Total exposure,
        ]   
        \addplot [mark=star] table[x=round, y=exposure]{\exposureaestwo}; 
        \label{plot_one}
    \end{axis}
    \begin{axis}[
        smooth,
        axis y line*=right,
        xlabel near ticks,
        ylabel near ticks,
        xmajorticks=false,
        y tick label style={font=\footnotesize},
        %ylabel=Worst net exposure,
        ylabel near ticks,
        ylabel style={font=\footnotesize},
        xlabel style={font=\footnotesize},
        width = 0.31\linewidth,
        legend pos=north east,
        legend style={font=\footnotesize},
        legend cell align=left,
        ]

        \addlegendimage{/pgfplots/refstyle=plot_one}
        \addlegendentry{$\sum$ exp}
   
        \addplot [mark=triangle, only marks] table[x=round, y=worst]{\exposureaestwo}; \label{plot_two}
        \addlegendimage{/pgfplots/refstyle=plot_two}
        \addlegendentry{Wne (\%)}
    \end{axis}   
\end{tikzpicture}
}%
\subfloat[Exposure for SPARX]{
\begin{tikzpicture}
    \begin{axis}[
        axis y line*=left,
        xlabel near ticks,
        ylabel near ticks,
        xlabel = Round,
        x tick label style={font=\footnotesize},
        y tick label style={rotate=45, anchor=east, font=\footnotesize},
        ylabel style={font=\footnotesize},
        xlabel style={font=\footnotesize},
        width =  0.31\linewidth,
        ymin = 0,
        %ylabel=Total exposure,
        %legend pos=north east,
        %legend style={font=\footnotesize},
        %legend cell align=left,
        ]   
        \addplot [mark=star] table[x=round, y=exposure]{\exposuresparx}; 
        \label{plot_three}
    \end{axis}
    \begin{axis}[
        smooth,
        ylabel style={anchor=south},
        axis y line*=right,
        xmajorticks=false,
        ylabel near ticks,
        y tick label style={font=\footnotesize},
        %ylabel=Worst net exposure (\%),
        ylabel near ticks,
        ylabel style={font=\footnotesize},
        xlabel style={font=\footnotesize},
        width = 0.31\linewidth,
        ymin = 30,
        ymax=50
        ] 
        \addplot [mark=triangle, only marks] table[x=round, y=worst]{\exposuresparx}; \label{plot_four}
    \end{axis}
    \label{fig:partfour}
\end{tikzpicture}
}%
\subfloat[Exposure for OMSP430\_ARRAY]{
\begin{tikzpicture}
    \begin{axis}[
        axis y line*=left,
        xlabel near ticks,
        ylabel near ticks,
        xlabel = Round,
        x tick label style={font=\footnotesize},
        y tick label style={rotate=45, anchor=east, font=\footnotesize},
        ylabel style={font=\footnotesize},
        xlabel style={font=\footnotesize},
        width =  0.31\linewidth,
        ymin = 1200,
        %ylabel=Total exposure,
        %legend pos=north east,
        %legend style={font=\footnotesize},
        %legend cell align=left,
        ]   
        \addplot [mark=star] table[x=round, y=exposure]{\exposurearray}; 
        \label{plot_three}
    \end{axis}
    \begin{axis}[
        smooth,
        ylabel style={anchor=south},
        axis y line*=right,
        xmajorticks=false,
        ylabel near ticks,
        y tick label style={font=\footnotesize},
        ylabel=Worst net exposure (\%),
        ylabel near ticks,
        ylabel style={font=\footnotesize},
        xlabel style={font=\footnotesize},
        width = 0.31\linewidth,
        ymin = 47,
        ymax= 53
        ] 
        \addplot [mark=triangle, only marks] table[x=round, y=worst]{\exposurearray}; \label{plot_six}
    \end{axis}
    \label{fig:partfour}
\end{tikzpicture}
}
% \end{adjustbox}
\caption{Evolution of TI and FSPFI strategies. \textbf{$\sum$exp} = summed exposure of all net assets. \textbf{Wne} = worst net exposure (\%).}
\label{fig:exposure}
\end{figure*}
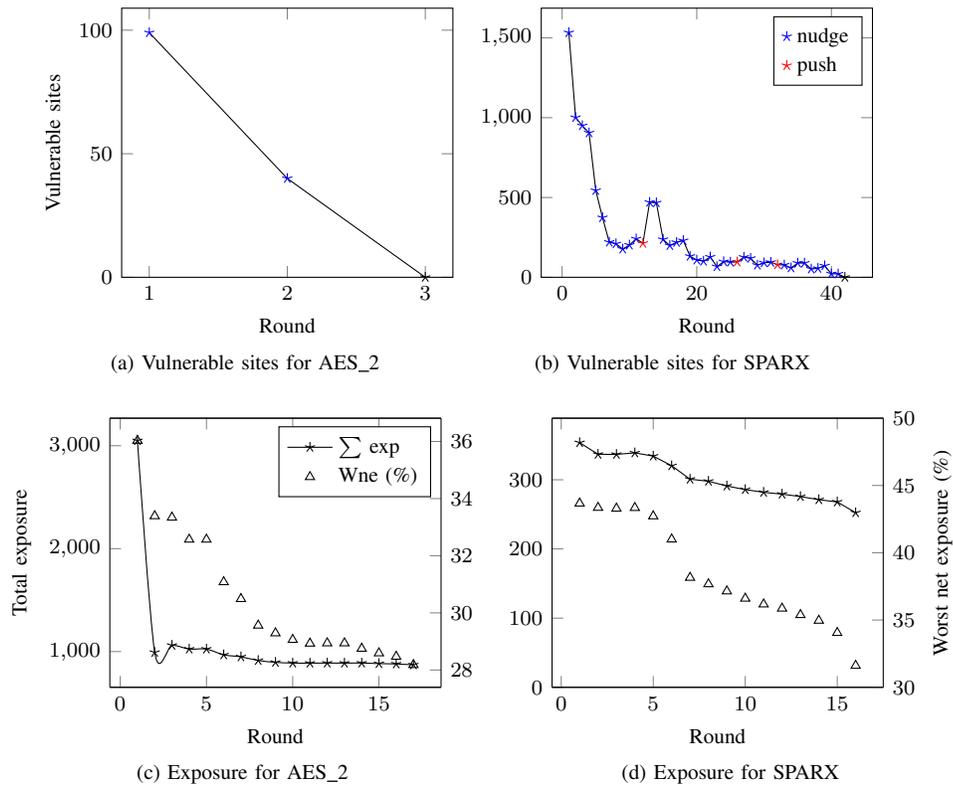

In order to understand how the TI and FSPFI strategies evolve through their rounds, a detailed breakdown of vulnerable sites and exposure is provided in Fig.~\ref{fig:exposure}. The chosen circuits are AES\_2 and SPARX because they have the highest FSPFI score and the highest number of cells touched during TI strategy, respectively. Notice that the \emph{pushes} in Fig.~\ref{fig:exposure}(b) cause the number of sites to increase temporarily, but this disturbance is needed for the next round of \emph{nudges} to succeed. Figures~\ref{fig:exposure}(c-d) highlight the importance of the two phases of the proposed FSPFI strategy: Phase 1 is highly effective in decreasing the total exposure while Phase 2 focuses on individual nets. This two-pronged approach is similar to WNS/TNS phases in timing optimization.

\pgfplotstableread[row sep=\\,col sep=&]{
    round & sites & style\\
    1 & 99 & nudge \\
    2 & 40 & nudge \\
    3 &  0 & none \\
}\vulnerableaestwo

\pgfplotstableread[row sep=\\,col sep=&]{
    round & sites & style\\
     1 & 1533 & nudge \\
     2 & 1001 & nudge \\
     3 &  950 & nudge \\
     4 &  905 & nudge \\
     5 &  544 & nudge \\
     6 &  374 & nudge \\
     7 &  220 & nudge \\
     8 &  210 & nudge \\
     9 &  177 & nudge \\
    10 &  202 & nudge \\
    11 &  241 & nudge \\
    12 &  213 & push \\
    13 &  470 & nudge \\
    14 &  467 & nudge \\
    15 &  237 & nudge \\
    16 &  199 & nudge \\
    17 &  219 & nudge \\
    18 &  230 & nudge \\
    19 &  132 & nudge \\
    20 &  108 & nudge \\
    21 &  101 & nudge \\
    22 &  127 & nudge \\
    23 &  67 & nudge \\
    24 &  98 & nudge \\
    25 &  95 & nudge \\
    26 &  96 & push \\
    27 &  126 & nudge \\
    28 &  119 & nudge \\
    29 &  76 & nudge \\
    30 &  91 & nudge \\
    31 &  94 & nudge \\
    32 &  79 & push \\
    33 &  77 & nudge \\
    34 &  59 & nudge \\
    35 &  89 & nudge \\
    36 &  89 & nudge \\
    37 &  52 & nudge \\
    38 &  56 & nudge \\
    39 &  72 & nudge \\
    40 &  22 & nudge \\
    41 &  21 & nudge \\
    42 &  0 & none \\
}\vulnerablesparx

\pgfplotstableread[row sep=\\,col sep=&]{
    round & sites & style\\
     1 & 7221 & nudge \\
     2 & 5259 & nudge \\
     3 &  3815 & nudge \\
     4 &  3440 & nudge \\
     5 &  3950 & nudge \\
     6 &  4117 & nudge \\
     7 &  3309 & nudge \\
     8 &  3183 & nudge \\
     9 &  3503 & nudge \\
    10 &  2632 & nudge \\
    11 &  3056 & nudge \\
    12 &  3113 & nudge \\
    13 &  2752 & push \\
    14 &  1600 & nudge \\
    15 &  1960 & nudge \\
    16 &  1896 & nudge \\
    17 &  1391 & nudge \\
    18 &  1009 & nudge \\
    19 &  913 & nudge \\
    20 &  1253 & nudge \\
    21 &  887 & nudge \\
    22 &  1141 & nudge \\
    23 &  825 & nudge \\
    24 &  1085 & nudge \\
    25 &  860 & nudge \\
    26 &  839 & push \\
    27 &  1126 & nudge \\
    28 &  1644 & nudge \\
    29 &  1460 & nudge \\
    30 &  1146 & nudge \\
    31 &  1210 & nudge \\
    32 &  1064 & push \\
    33 &  1036 & nudge \\
    34 &  939 & nudge \\
    35 &  997 & nudge \\
    36 &  944 & nudge \\
    37 &  1006 & push \\
    38 &  846 & nudge \\
    39 &  672 & nudge \\
    40 &  438 & nudge \\
    41 &  303 & nudge \\
    42 &  274 & nudge \\
    43 &  323 & nudge \\
    44 &  293 & nudge \\
    45 &  262 & nudge \\
    46 &  400 & nudge \\
    47 &  294 & nudge \\
    48 &  297 & push \\
    49 &  929 & nudge \\
    50 &  756 & nudge \\
    51 &  695 & nudge \\
    52 &  330 & nudge \\
    53 &  288 & nudge \\
    54 &  261 & nudge \\
    55 &  200 & nudge \\
    56 &  165 & nudge \\
    57 &  194 & nudge \\
    58 &  157 & nudge \\
    59 &  185 & nudge \\
    60 &  172 & nudge \\
    61 &  164 & push \\
    62 &  141 & nudge \\
    63 &  177 & nudge \\
    64 &  107 & nudge \\
    65 &  118 & nudge \\
    66 &  100 & nudge \\
    67 &  55 & nudge \\
    68 &  91 & nudge \\
    69 &  108 & nudge \\
    70 &  91 & push \\
    71 &  185 & nudge \\
    72 &  123 & nudge \\
    73 &  110 & nudge \\
    74 &  108 & nudge \\
    75 &  62 & nudge \\
}\vulnerablearray

\pgfplotstableread[row sep=\\,col sep=&]{
    round & exposure & worst \\
    1 & 3049.65	& 36.03 \\
    2 &  991.73 & 33.39 \\
    3 & 1060.84 & 33.34 \\
    4 & 1024.39 & 32.57 \\
    5 & 1024.37 & 32.57 \\
    6 &  968.27 & 31.08 \\    
    7 &  948.86 & 30.49 \\
    8 &  913.28 & 29.56 \\
    9 &  895.35 & 29.29 \\
    10 &  888.22 & 29.06 \\
    11 &  887.65 & 28.93 \\
    12 &  888.33 & 28.94 \\
    13 &  887.94 & 28.94 \\
    14 &  888.67 & 28.75 \\
    15 &  883.13 & 28.59 \\
    16 &  879.62 & 28.47 \\
    17 &  870.81 & 28.18 \\
}\exposureaestwo

\pgfplotstableread[row sep=\\,col sep=&]{
    round & exposure & worst \\
    1 & 353.64 & 43.67 \\
    2 & 336.82 & 43.35 \\
    3 & 336.71 & 43.29 \\
    4 & 338.85 & 43.34 \\
    5 & 334.22 & 42.72 \\
    6 & 320.23 & 41.00 \\    
    7 & 300.84 & 38.15 \\
    8 & 297.83 & 37.67 \\
    9 & 291.01 & 37.14 \\
   10 & 285.80 & 36.60 \\
   11 & 281.84 & 36.17 \\
   12 & 279.22 & 35.86 \\
   13 & 275.71 & 35.38 \\
   14 & 271.21 & 34.96 \\
   15 & 267.92 & 34.05 \\
   16 & 252.40 & 31.61 \\
}\exposuresparx

\pgfplotstableread[row sep=\\,col sep=&]{
    round & exposure & worst \\
    1 & 1307.49 & 51.89 \\
    2 & 1306.62 & 51.56 \\
    3 & 1299.57 & 50.93 \\
    4 & 1295.11 & 50.35 \\
    5 & 1291.30 & 49.97 \\
    6 & 1269.89 & 49.27 \\    
    7 & 1253.41 & 48.74 \\
    8 & 1253.41 & 48.74 \\
    9 & 1253.41 & 48.74 \\
   10 & 1253.41 & 48.74 \\
   11 & 1253.41 & 48.74 \\
   12 & 1253.41 & 48.74 \\
   13 & 1253.41 & 48.74 \\
   14 & 1253.41 & 48.74 \\
   15 & 1253.40 & 48.74 \\
   16 & 1252.51 & 48.62 \\
   17 & 1252.51 & 48.65 \\
   18 & 1251.54 & 48.61 \\
   19 & 1249.97 & 48.50 \\
   20 & 1249.97 & 48.50 \\
   21 & 1237.91 & 48.01 \\
   22 & 1231.11 & 47.81 \\
   23 & 1231.09 & 47.81 \\
   24 & 1231.09 & 47.81 \\
}\exposurearray

%Reported area values are for die area (i.e., floorplan bounding box) and not for the cell or net area. Power is total power, i.e., the sum of dynamic and static power components. 

%Implementation assumes the design area has been iterated over and the smallest routable/closable area has been found. Tool load time and library read times are not considered. In practice, the implementation requires a handful of attempts. Execution times do not include scoring, which is executed with a third-party set of scripts and binaries provided by the contest organizers.

\pgfplotstableread[row sep=\\,col sep=&,columns/Benchmark/.style={string type}]{
Benchmark & winner \\
    AES\_1        & 0.447645 \\
    AES\_2        & 0.425056 \\
    AES\_3        & 0.473199 \\
    CAMELLIA      & 0.398203 \\
    CAST          & 0.412035 \\
    MISTY         & 0.418306 \\
    OMSP430\_1    & 0.406426 \\
    OMSP430\_2    & 0.464010 \\
    PRESENT       & 0.359781 \\
    SEED          & 0.416061 \\
    SPARX         & 0.397067 \\
    TDEA          & 0.459273 \\
}\desmetricwinner

\pgfplotstableread[row sep=\\,col sep=&]{
    Benchmark     & overall \\
    AES\_1        & 0.14563 \\
    AES\_2        & 0.19036 \\ 
    AES\_3        & 0.15694 \\
    CAMELLIA      & 0.14726 \\
    CAST          & 0.15196 \\
    MISTY         & 0.13824 \\
    OMSP430\_1    & 0.16465 \\
    OMSP430\_2    & 0.21135 \\
    PRESENT       & 0.12817 \\
    SEED          & 0.14295 \\
    SPARX         & 0.13542 \\
    TDEA          & 0.16689 \\
}\overallscore

We compare, in Tab.~\ref{tab:ispdwinners}, area and power results versus those obtained by the team that got the first-place award at the contest. It is evident that our IMP strategy compromises \textbf{no area}; on average, our results shrink area by 9.22\%. Regarding power, our results show an average reduction of 2.78\%. Regarding the DES score (Eq.~\ref{eq1}) as a whole, the contest winners obtained an average score of 0.42309, whereas our combined strategies led to an improved average DES score of 0.39842 (lower numbers are better).

\begin{table*}
\caption{Area and power comparison vs. the contest winner.}
\label{tab:ispdwinners}
\begin{threeparttable}[htbp]
{\setlength{\tabcolsep}{10pt}
\renewcommand{\arraystretch}{1.3}
\renewcommand{\TPTminimum}{\linewidth}
\makebox[\linewidth]{%
\begin{tabular}{lllclcr}
\hline
\multicolumn{1}{c}{\textbf{Benchmark}} & \multicolumn{2}{c}{\textbf{Area ($\si{\micro\meter}^2$)}} & \multicolumn{2}{c}{\textbf{Static Power ($mW$)}}& \multicolumn{2}{c}{\textbf{Total Power ($mW$)}}   \\ 
 & \multicolumn{1}{c}{\textbf{ISPDw}} & \multicolumn{1}{c}{\textbf{TW}} &  \multicolumn{1}{c}{\textbf{ISPDw}} & \multicolumn{1}{c}{\textbf{TW}} &  \multicolumn{1}{c}{\textbf{ISPDw}} & \multicolumn{1}{c}{\textbf{TW}} \\ \hline
\rowcolor[HTML]{EFEFEF} AES\_1 &        40813.6 & 37039.0 (-10.1\%) & 0.737 & 0.705 (-4.5\%) & 66.14 & 65.94 (-0.3\%) \\ 
AES\_2 &        40813.6 & 36881.1 (-10.6\%) & 0.742 & 0.703 (-5.5\%) & 62.13 & 60.41 (-2.8\%) \\
\rowcolor[HTML]{EFEFEF} AES\_3 &        40149.9 & 36881.1 (-8.8\%) & 0.750 & 0.697 (-7.6\%) & 60.46 & 61.33 (+1.4\%) \\ 
                        CAMELLIA &      11071.7 & 10137.6 (-9.2\%) & 0.147 & 0.145 (-1.3\%) & 1.71 & 1.69 (-1.1\%) \\
\rowcolor[HTML]{EFEFEF} CAST &          17954.3 & 16132.3 (-11.2\%) & 0.263 & 0.254 (-3.5\%) & 4.81 & 4.55 (-5.7\%) \\
                        MISTY &         14346.1 & 12424.8 (-15.4\%) & 0.202 & 0.188 (-7.4\%) & 3.41 & 3.15 (-8.2\%) \\
\rowcolor[HTML]{EFEFEF} OMSP430\_1 &    10376.8 &  9965.5 (-4.1\%) & 0.108 & 0.111 (+2.7\%) & 0.40 & 0.39 (-2.5\%) \\
                        OMSP430\_2 &    11787.4 & 11236.0 (-4.9\%) & 0.128 & 0.129 (+0.8\%) & 1.15 & 1.13 (-1.7\%) \\ 
\rowcolor[HTML]{EFEFEF} PRESENT &        2409.5 &  2118.6 (-13.7\%) & 0.020 & 0.020 (+0.0\%) & 0.32 & 0.31 (-3.2\%)\\
                        SEED &          17954.3 & 16170.6 (-11.0\%) & 0.265 & 0.254 (-4.3\%) & 4.83 & 4.45 (-8.5\%) \\
\rowcolor[HTML]{EFEFEF} SPARX &         21911.1 & 20141.9 (-8.7\%) & 0.264 & 0.261 (-1.1\%) & 2.26 & 2.24 (-0.8\%)\\
                        TDEA &           4455.6 &  4325.7 (-3.0\%) & 0.046 & 0.046 (+0.0\%) & 1.45 & 1.45 (+0.0\%) \\ \hline
\end{tabular}
}
 \begin{tablenotes}
 \centering
    \item \small \textbf{ISPDw} = winner of the ISPD 2022 contest. \textbf{TW} = this work.
\end{tablenotes}
}
\end{threeparttable}
\end{table*}

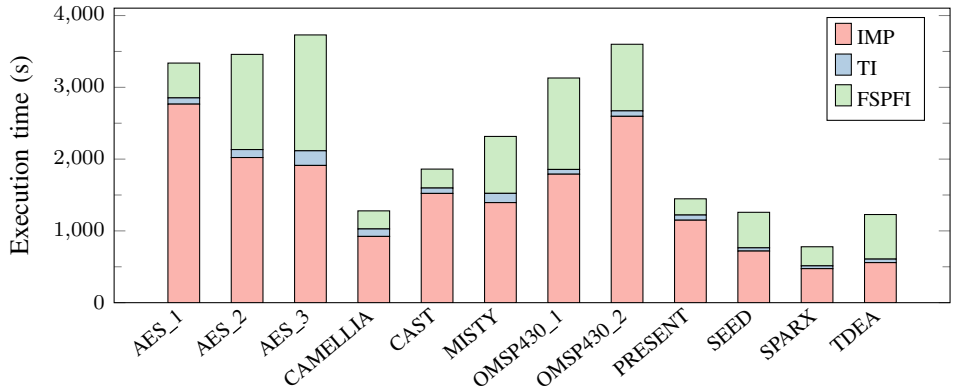
\begin{figure*}[htbp]
\centering
\begin{tikzpicture}
    \begin{axis}[
        ybar stacked,
        symbolic x coords={AES\_1, AES\_2, AES\_3, CAMELLIA, CAST, MISTY, OMSP430\_1, OMSP430\_2, PRESENT, SEED, SPARX, TDEA, OMSP430\_ARRAY},
        xtick=data,
        xtick pos = bottom,
        x tick label style={rotate=40, anchor=east,font=\footnotesize, yshift=-2pt},
        y tick label style={font=\footnotesize},
        ymode=log,
        width = 0.7\textwidth,
        height = 5.5cm,
        minor y tick num = 1,
        ymin = 0,
        ylabel={Execution time (s, log scale)},
        legend pos=north west,
        legend style={font=\footnotesize},
        legend cell align=left,
        bar width=12pt,
        ]
        \addplot[fill=mycolor1] table[x=Benchmark,y=imp]{\exectime};
        \addplot[fill=mycolor2] table[x=Benchmark,y=ti_strat]{\exectime};
        \addplot[fill=mycolor3] table[x=Benchmark,y=fsp_strat]{\exectime};
        
        \addlegendentry{IMP}
        \addlegendentry{TI}
        \addlegendentry{FSPFI}
    \end{axis}
\end{tikzpicture}
% \end{adjustbox}
\caption{Execution times for IMP, TI, and FSPFI strategies.}
\label{fig:exec_times}
\end{figure*}

The execution times of the three strategies are given in Fig.~\ref{fig:exec_times}. Note that the implementation strategy is considerably more demanding than our security-aware strategies. This is remarkable since the implementation flow benefits significantly from multithreading while the TI and FSPFI strategies do not due to their ECO-like approach. Also note that the reported execution times do not include any library or design loading, nor does it include scoring that is performed by an external binary provided by the contest organizers. 

\pgfplotstableread[row sep=\\,col sep=&]{
    Benchmark   & imp & ti_strat & fsp_strat \\
    AES\_1        & 2766 &  87 &  485 \\  
    AES\_2        & 2022 & 111 & 1326 \\
    AES\_3        & 1912 & 204 & 1614 \\
    CAMELLIA      &  923 & 104 &  252 \\
    CAST          & 1522 &  75 &  264 \\
    MISTY         & 1394 & 129 &  793 \\ 
    OMSP430\_1    & 1790 &  66 & 1274 \\
    OMSP430\_2    & 2596 &  76 &  928 \\
    PRESENT       & 1149 &  72 &  226 \\
    SEED          &  719 &  45 &  495 \\
    SPARX         &  474 &  39 &  266 \\
    TDEA          &  558 &  49 &  621 \\
    OMSP430\_ARRAY & 7488 &  89  & 4725 \\ 
    %total nets touched in B round is 4219
    }\exectime 

\subsection{Scalability Analysis}
\label{sec:scalability}

For demonstrating scalability of our approaches, we now consider a larger design that contains 64 instances of the MSP430 microcontroller, forming an array of processors that operate in lockstep. Only one of the processors (cpu1) is considered security critical and contains cell asssets and net assets. The array contains 360k cells, 410k nets, and achieves timing closure at 125MHz.

Contrary to the other benchmarks, the array we devised is hierarchical.  One of our findings is that flattened blocks tend to have TI vulnerable sites on the periphery of the floorplan, whereas a large hierarchical block will have vulnerable sites also at the inner boundaries of the hierarchy. Nevertheless, our strategies executed successfully on this large design.

The first result we would like to highlight is given in panels (c) and (f) of Fig.~\ref{fig:exposure}. Notice that the evolution of the TI strategy is very similar to that of SPARX design and convergence is reached after 76 rounds. For the FSPFI strategy, the round by round behavior is smooth, consistently moving the design towards a more secure layout. The process ends at round 24.

\begin{figure}[htbp]
    \centering
    \begin{tikzpicture}[spy using outlines={circle,gray,magnification=3,size=1.8cm, connect spies}]
            \node {\includegraphics[width=0.60\columnwidth, trim=0cm 0cm 0cm 0cm]{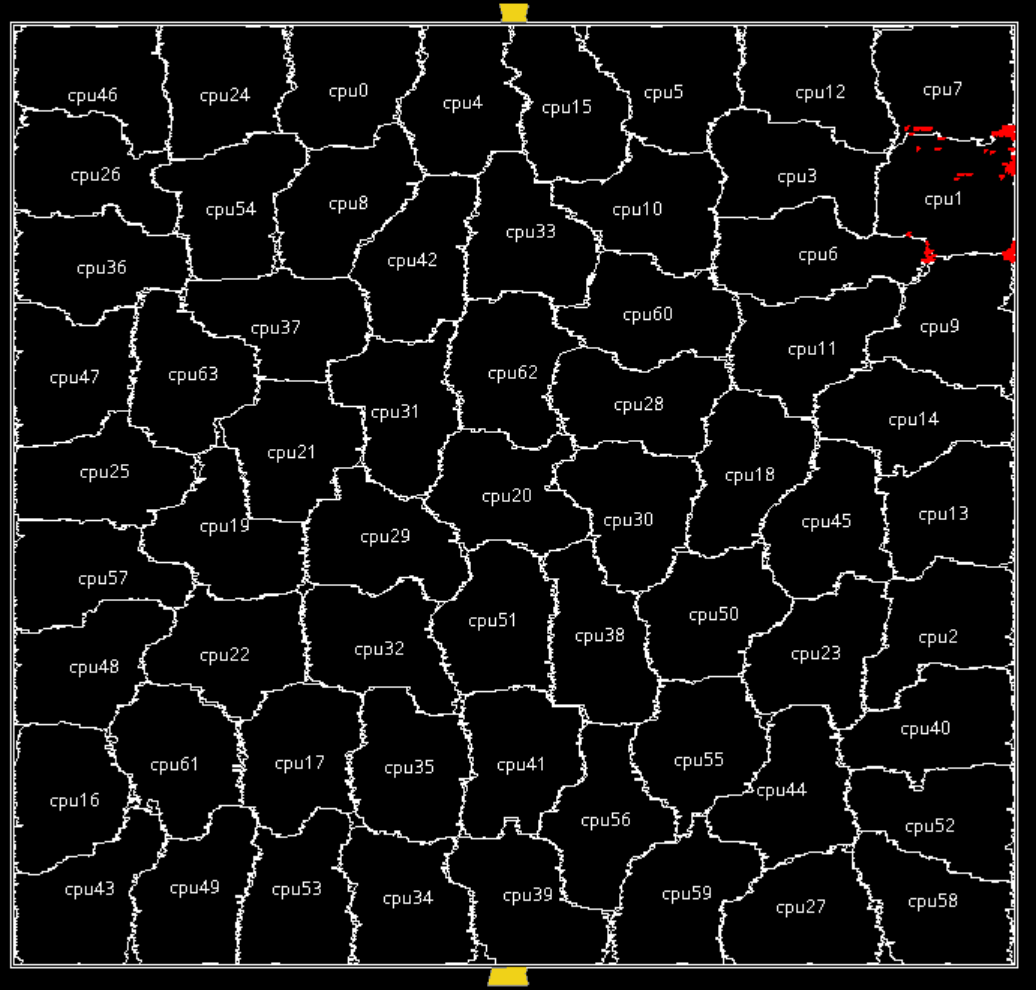}};
            \spy on (2.18,1.33) in node [right] at (2.8,1.5);
    \end{tikzpicture}
    \caption{Amoeba floorplan view of the processor array. A total of 64 processors are instantiated, only `cpu1' is considered security critical.}
    \label{fig:array}
\end{figure}

The amoeba view of the processor array floorplan can be seen in Fig.~\ref{fig:array}. Note that the floorplan measures 840um x 788 um and that the security-critical processor is located on the top right. The zoomed in area shows the vulnerable sites that are located right at the boundary between cpu1 and its neighbors. Figure~\ref{fig:array} is a representation of the design right after IMP was executed, therefore vulnerable sites are still present.

\begin{figure}[htbp]
    \centering
    \begin{tikzpicture}[]
            \node {\includegraphics[width=0.74\columnwidth, trim=0cm 0cm 0cm 0cm]{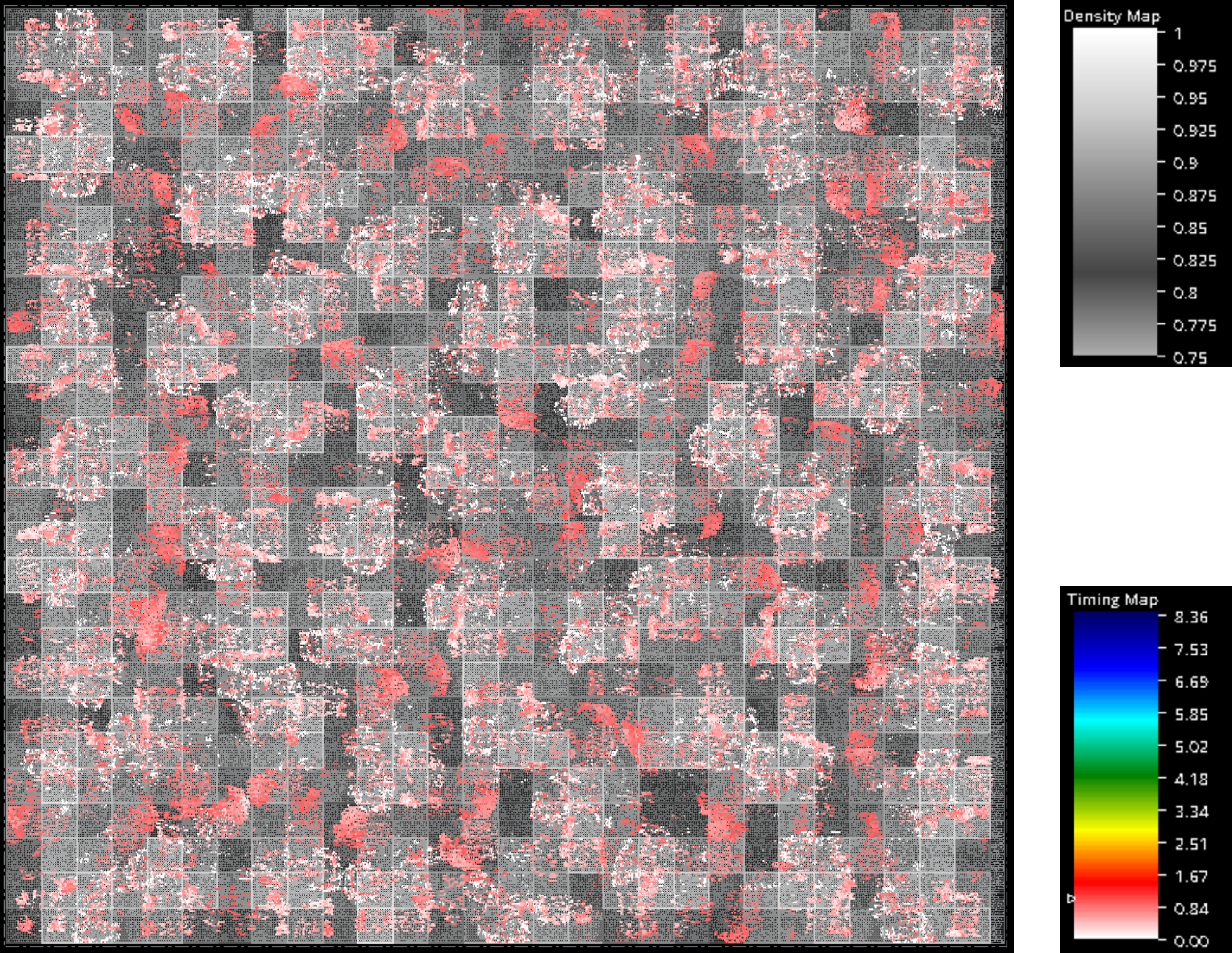}};
            
    \end{tikzpicture}
    \caption{Physical floorplan view of the processor array, overlaying a density map (gray scale) and a timing map (color). The design measures 840um x 788um. }
    \label{fig:arraymaps}
\end{figure}

Finally, Fig.~\ref{fig:arraymaps} shows two maps overlayed on top of the finalized design, i.e., after TI and FSPFI strategies were executed. The coarse-grained grayscale rectangles show a placement density map, whereas the fine-grained colored rectangles show a timing map. The timing map is filtered such that it only shows cells on paths that have between 0ps and 1000ps of positive slack. The main insight from this analysis is that our approaches, even when applied to an hierarchical large design, do not compromise PPA. Visually, Fig.~\ref{fig:arraymaps} shows no distinctive pattern near the security-critical module, i.e., cpu1. There are no exacerbated critical paths or overly congested placement regions.

Execution times for the processor array are shown in Fig.~\ref{fig:exec_times}. It is clear that the FSPFI strategy is much more demanding than the TI strategy. It is also clear that the FSPFI strategy is slightly less complex than the IMP strategy. Notably, the execution times of TI and FSPFI strategies are kept under control given the ECO nature of both.

After implementation, the processor array has a cell area of $586,567 um^2$, a slack of $0 ps$, a total power consumption of 78.20$mW$, and a DES score of 0.45. The TI strategy increases cell area by 0\% and power by 0.0001\%. The FSPFI strategy decreases area by 0.004\% and increases power by 0.2\%. The DES score increases by 0.1\%. In general, we emphasize that the proposed strategies remain effective even for large hierarchical designs. The PPA overheads we report are either negligible or heuristic artifacts.

\subsection{Sensitivity Analysis}
\label{sec:sensitivity}

Our proposed strategies have two main parameters (i.e., $STUCK\_MAX$ and $NETS\_PER\_RD$) that influence the execution time and the QoR of the experiments. In order to evaluate how sensitive the methodology is to these parameters, we perform two experiments. First, we evaluate different values for $STUCK\_MAX$ on the SPARX design -- SPARX was chosen according to the results shown in Tab.~\ref{tab:completeresults} which reveal that SPARX has the hardest TI scenario, requiring the highest number of iterations.

{
\color{blue}
\begin{table}[tb]
\caption{TI scores, DES scores, and execution times with respect to $STUCK\_MAX$ parameter. The benchmark considered is SPARX.}
\label{tab:stuckmax}
\centering
\renewcommand{\arraystretch}{1.3}
\begin{tabular}{ccccc}
\hline
\textbf{$STUCK\_MAX$}   & \textbf{Rounds (N/P)} & \textbf{TI} & \textbf{DES} & \textbf{Exec. time (s)} \\ \hline
\rowcolor[HTML]{EFEFEF} 1           & 79 (0/79)      & 0       & 0.37514   & 43 \\
2           & 71 (10/61)     & 0        & 0.37479 & 39 \\
\rowcolor[HTML]{EFEFEF}3 (adopted) & 41 (38/3)      & 0        & 0.37477 & 36 \\
4           & 54 (49/5)      & 0        & 0.37477 & 35 \\
\rowcolor[HTML]{EFEFEF}5           & 50 (48/2)      & 0        & 0.37478 & 34 \\ \hline
\end{tabular}
\end{table}
}

Results for the sensitivity analysis are shown in Tab.~\ref{tab:stuckmax} and they reveal that any value assigned to $STUCK\_MAX$ will yield a TI score of zero, indicating that the security vulnerability can be fully addressed. Execution times reveal no significant difference, whereas the DES score reveals that doing only pushes can be detrimental ($STUCK\_MAX$ = 1 means nudges are never performed). Finally, the adopted value ($STUCK\_MAX$ = 3) reveals a good balance that also reduces the total number of rounds required.

Next, we evaluate different values for $NETS\_PER\_RD$ on the AES\_3 design -- AES\_3 was chosen according to the results shown in Tab.~\ref{tab:completeresults} which reveal that AES\_3 has the hardest FSPFI scenario, requiring the highest number of nets to be rerouted during ECO. Results are shown in Tab.~\ref{tab:netsperrd}.

\begin{table*}[tb]
\caption{Details of the FSPFI strategy with respect to the $NETS\_PER\_RD$ parameter for the AES\_3 benchmark.}
\label{tab:netsperrd}
\renewcommand{\arraystretch}{1.3}
\centering
\begin{tabular}{lccc|ccc|cr}
\hline
\textbf{$NETS\_PER\_RD$} & \textbf{R (A)} & \textbf{ECO'd (A)} & \textbf{Exec. time (s)} & \textbf{R (B)} & \textbf{ECO'd (B)} & \textbf{Exec. time (s)} & \textbf{FSPFI norm.} & \textbf{DRCs?}  \\ \hline
\rowcolor[HTML]{EFEFEF} 25 & 4 &  72 & 57 & 26 & 342 & 859 & 0.984 & NO  \\
75 & 5 & 116 & 96 & 13 & 368 & 571 & 0.999 & NO  \\
\rowcolor[HTML]{EFEFEF} 85 & 3 & 116 & 60 & 13 & 495 & 708 & 0.990 & NO  \\
94* (baseline) & 4 & 125 & 81 & 16 & 582 & 810 & 1 & NO  \\
\rowcolor[HTML]{EFEFEF} 105 & 3 & 133 & 63 & 2 & 155 & 138 & 1.065 & YES  \\
115 & 4 & 142 & 85 & 9 & 458 & 497 & 0.990 & YES  \\
\rowcolor[HTML]{EFEFEF} 125 & 3 & 154 & 64 & 12 & 568 & 826 & 0.962 & NO  \\
375 & 3 & 408 & 76 & 11 & 602 & 777 & 0.989 & YES \\ \hline
\end{tabular}
\end{table*}

From the data presented in Tab.~\ref{tab:netsperrd}, it is possible to infer some high-level insights: (i) execution time is dominated by phase B of the strategy; (ii) DRC violations that are not resolvable tend to appear with higher values of the parameter but their appearance is \textbf{heuristic in nature}; (iii) the lower the parameter, the higher the number of rounds executed by the FSPFI strategy; (iv) lower parameter values can yield better FSPFI scores by small margins (less than 2\% improvement); (v) The number of nets touched is proportional to the parameter and the number of rounds.

Note that the table contains somewhat extreme values for $NETS\_PER\_RD$ (i.e., 25 and 375). Also note that the case when $NETS\_PER\_RD$=105 reveals a scenario where a DRC was generated on the second iteration of phase B, causing an early abort and a very poor FSPFI score. In other words, this indicates the parameter has a heuristic outcome. While not shown on the table, a value of 106 does not suffer from the same issue, again highlighting the heuristic nature. One other aspect that is not captured by the data in Tab.~\ref{tab:netsperrd} is that many calls to an external binary that performs scoring are needed, especially for the case when $NETS\_PER\_RD$=25. These calls are time consuming as they require significant exchange of data between Innovus and the external binary. This could be mitigated if scoring was fully integrated within the physical synthesis flow.

In summary, selecting a value for the $NETS\_PER\_RD$ parameter is not entirely trivial as it has a heuristic nature to it (technically, the heuristic nature is inherited from the ECO engine that does the rerouting after every round). On the other hand, the $STUCK\_MAX$ parameter is much more constrained in its influence on QoR.

\section{Discussion} \label{sec_discusstion}

%In \cite{placement_strategies}, the authors propose placement refinement algorithms that leverage clustering and heuristic cell movement to reduce vulnerable placement sites with minimal impact on performance and wirelength. While more efficient than exhaustive approaches, the method risks reintroducing vulnerable regions during subsequent optimization steps if not applied in the late stages of the design flow.

A key outcome of our study is the demonstration that it is possible to achieve robust security closure without compromising the design’s PPA. While most existing approaches introduce noticeable PPA penalties as a trade-off for improved security, our results challenge this convention entirely. The SALSy methodology~\cite{salsy} results in leakage power increases ranging from 1.6\% to 18.8\% due to buffer insertion. SALSy also yields dynamic power increases that are relatively small, typically under 3\%. ASSURER~\cite{ASSURER} reports power increases in the range of 2\% to 14\% (total power) due to buffer insertion. GDSII-Guard~\cite{gdsguard} does not have an FSPFI component, it only protects a design against the TI threat. GDSII-Guard results reveal several shortcomings: it does not achieve 100\% TI mitigation like our work does, it \textbf{does not} close timing, and it also creates non-negligible power overheads (as high as 18\% for the MISTY design).

Furthermore, in prior work such as ASSURER~\cite{ASSURER}, results often display increased power consumption due to a reliance on global design modifications. In contrast, we observe that, in most cases, our approach not only avoids such overheads but also leads to a reduction in total power. This makes it particularly suitable for designs operating under tight PPA constraints. Prior methods~\cite{placement_strategies,gdsguard} also sometimes focus solely on placement-level interventions. While relatively efficient, these methods are inherently limited in scope. Without accounting for routing-level vulnerabilities, they fall short of achieving the holistic security improvements enabled by our combined approach. In contrast, our flow leverages combined optimization across both P\&R tasks. This allows for a more comprehensive and fine-grained security closure, improving resilience against both FSPFI and TI. It also allows us to swiftly address other attack vectors while keeping the same methodology in place.

One significant threat vector to consider would be power side-channel analysis (SCA), where an adversary collects power traces in order to establish a correlation with a secret (i.e., a key). SCA attacks succeed because the design's power consumption is modulated by the inputs provided to it. If those inputs are secrets, then a power trace is secret-dependent (to an extent). While our strategies do not directly address SCA, they do not contribute to the success of an attack either. The act of moving cells or rerouting cells does not change the fact that there will be gates that will deal with inputs that are secrets. However, it is important to mention that our flow is compatible with diverse SCA protection mechanisms proposed in the literature: The foundational work on constant-power cell style described in~\cite{constant_power} could be considered in conjunction with our TI strategy with relative ease; the logic synthesis aspects of \cite{synthesis_sca} are fully compatible with our work; masking logic~\cite{firstmasking}, in general, is also entirely compatible with our strategies.

Equally important is the scalability of our approach. Thanks to its ECO-like nature, where changes are confined to \textbf{localized placement and routing adjustments}, the methodology is inherently scalable and remains effective across a wide range of design sizes. From small modules with a few hundred gates to full-scale industrial designs containing hundreds of millions of cells, the flow operates efficiently without requiring global restructuring.

A comparative analysis of existing security closure techniques, summarized in Table~\ref{tab:comparison}, further reinforces the advantages of our approach. While most prior methods achieve some level of security improvement, they often do so at the expense of increased power, area, or timing overhead, compromising practical deployment and adoption. Techniques such as logic locking or multiobjective placement optimization frequently change the problem complexity significantly: logic locking requires the insertion of a tamper-proof memory, while multiobjective optimization may require design-specific tuning. We believe both of these factors limit scalability and applicability to commercial flows. 

Furthermore, and notably, some frameworks remain validated only on open source PDKs, casting doubt on their industrial relevance. In contrast, our flow maintains compatibility with standard design practices while addressing diverse security threats. The minimal PPA disruption and localized implementation make it especially suitable for modern ASIC design environments, where both security and efficiency are critical. This positions our methodology as not only technically effective but also practically viable for widespread adoption.

\begin{table*}[htbp]
\centering
\caption{Comparative Overview of Security Closure Approaches} \label{tab:comparison}
\begin{tabular}{lllcll}
\toprule
\textbf{Approach}    & \textbf{Key Technique}                                                                          & \multicolumn{1}{c}{\textbf{Security Focus}}                                     & \multicolumn{1}{c}{\textbf{PPA Impact}}                                             & \multicolumn{1}{c}{\textbf{Scalability}}                                    & \multicolumn{1}{c}{\textbf{Notable Limitations}}                                                                               \\ \midrule
\rowcolor[HTML]{EFEFEF} 
DEFence \cite{DEFend}{} & \begin{tabular}[c]{@{}l@{}}Placement, routing, shielding\end{tabular} & TI, FSP, CT & \resizebox{4mm}{3mm}{\Neutrey} & Limited & \begin{tabular}[c]{@{}l@{}}  Performs logic restructuring to solve TI  \\  vulnerable zones; relies on shielding \\  using additional metal shapes\end{tabular} \\

ASSURER \cite{ASSURER} & \begin{tabular}[c]{@{}l@{}}Placement, ECO routing\end{tabular} & TI, FSP        & \resizebox{4mm}{3mm}{\Neutrey} & Moderate & \begin{tabular}[c]{@{}l@{}}Susceptible to refinement-reversal attacks \\ that may enable malicious insertions
\end{tabular} \\

\rowcolor[HTML]{EFEFEF} 
GDSII-Guard \cite{gdsguard} & \begin{tabular}[c]{@{}l@{}}Placement,\\ multi-objective optimization\end{tabular} & TI & \resizebox{4mm}{3mm}{\color[HTML]{fe0000}\Sadey} & Limited & \begin{tabular}[c]{@{}l@{}}Custom solutions for each design,\\ limited scalability and timing degradation\end{tabular} \\

TroLLoc \cite{TroLLoc} & \begin{tabular}[c]{@{}l@{}}Logic locking, placement \end{tabular} & TI & \resizebox{4mm}{3mm}{\color[HTML]{fe0000}\Sadey} & Limited & \begin{tabular}[c]{@{}l@{}} Involves significant cost, complexity, \\ and tight floorplanning constraints \end{tabular} \\

\rowcolor[HTML]{EFEFEF} 
TroMUX \cite{TroMUX} & \begin{tabular}[c]{@{}l@{}}Logic locking, placement \end{tabular} & TI & \resizebox{4mm}{3mm}{\color[HTML]{fe0000}\Sadey} & Limited & \begin{tabular}[c]{@{}l@{}}Relies on tamper-proof memory; \\ adds overheads and integration challenges\end{tabular} \\

SALSy\cite{salsy} & \begin{tabular}[c]{@{}l@{}}Placement, CTS, routing, \\ ECO routing, buffer insertion \end{tabular} & TI, FSP, FI & \resizebox{4mm}{3mm}{\Neutrey} & High & \begin{tabular}[c]{@{}l@{}}Increases power consumption due to \\ buffer insertion in sensitive layout regions\end{tabular} \\

\rowcolor[HTML]{EFEFEF} 
Placement-only \cite{placement_strategies} & \begin{tabular}[c]{@{}l@{}}Heuristic placement-level \\ refinement\end{tabular} & TI & \resizebox{4mm}{3mm}{\Neutrey} & Limited & \begin{tabular}[c]{@{}l@{}}Ignores routing-level vulnerabilities,\\  late-stage use may reintroduce weaknesses\end{tabular} \\

\textbf{This Work}   & \begin{tabular}[c]{@{}l@{}}Joint placement + routing \\ enhancements (ECO-like)\end{tabular} & TI, FSP, FI & \resizebox{4mm}{3mm}{\color[HTML]{009900}\Smiley} & High & -- (No notable limitations) \\ 
\bottomrule
\end{tabular}

\vspace{0.5em}
\begin{minipage}{\textwidth}
\centering
\small \textbf{TI} = Trojan insertion. \textbf{FSP} = Front-side probing. \textbf{CT} = Cross talk. \textbf{FI} = Fault injection.
\end{minipage}
\end{table*}

\section{Conclusion} \label{sec:conclusion}

In this paper, we presented a zero-overhead, security-aware ASIC design flow that integrates seamlessly with a commercial physical synthesis toolchain. Our methodology stands apart from prior approaches by delivering robust security closure without incurring penalties in PPA -- a key barrier in many existing solutions. The proposed flow effectively mitigates critical hardware security threats, including HTs and FSP/FI, while maintaining full DRC compliance.

%Through a combination of targeted placement and routing-level enhancements, our approach enables fine-grained security improvements that are both scalable and practical for real-world designs. 

Experimental results demonstrate that our method not only preserves design integrity but also achieves superior area and power efficiency compared to state-of-the-art techniques, making it highly suitable for deployment in PPA-constrained environments.
By open-sourcing the methodology and design databases, we aim to contribute to the hardware security community and promote the adoption of secure, efficient IC design practices.

\section*{Acknowledgment}
The authors would like to thank the ISPD'22 organizers for providing an offline version of the scoring scripts/engines.

% \vfill\eject
% \clearpage

\bibliographystyle{IEEEtran}
\bibliography{ref}

 \vfill\eject

\begin{IEEEbiography}[{\includegraphics[width=1in,height=1.25in,clip,keepaspectratio]{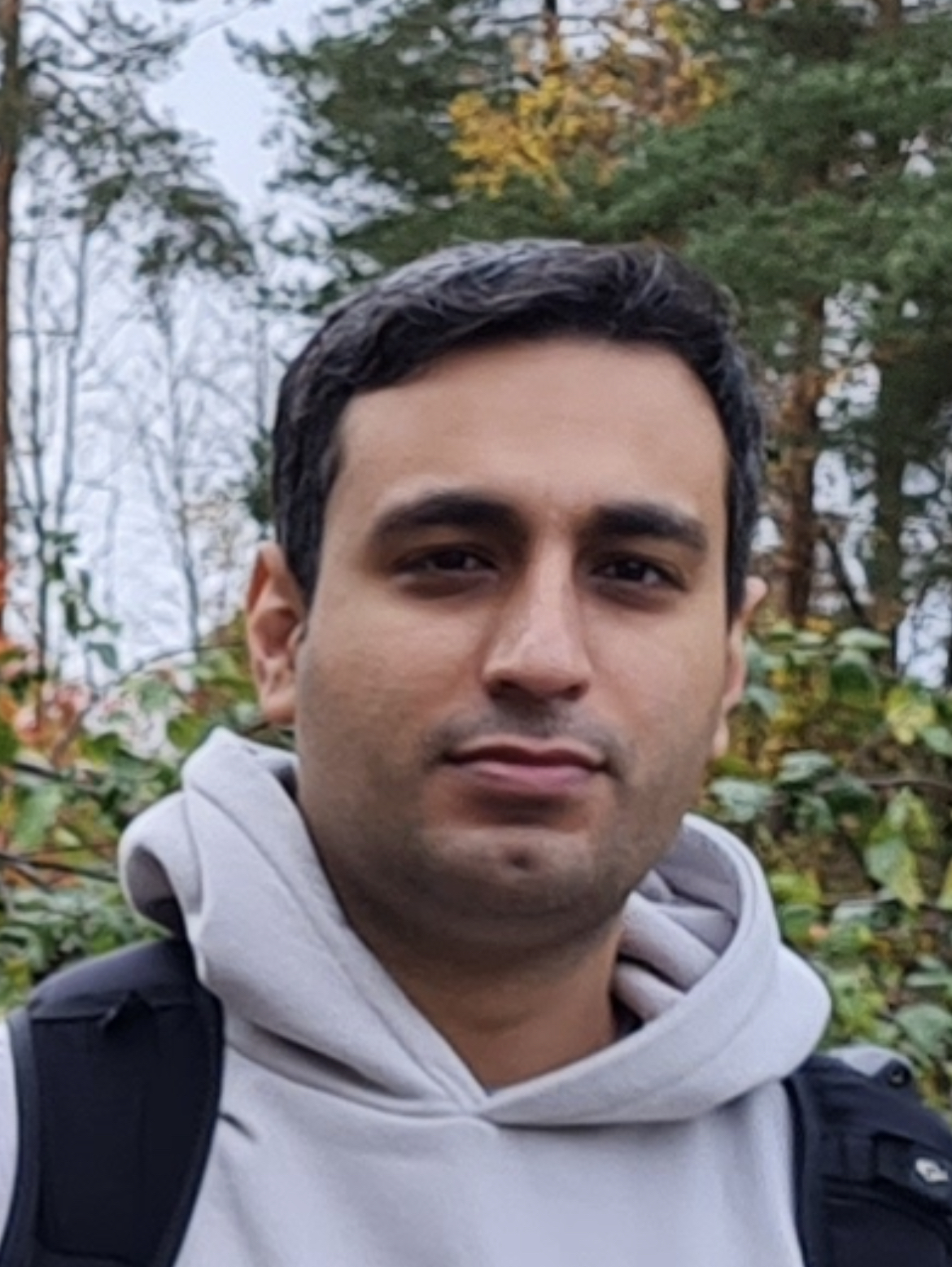}}]{Mohammad Eslami}
received his M.S. degree in Computer Engineering from Shahid Bahonar University of Kerman, Kerman, Iran, in 2018, and his Ph.D. degree from Tallinn University of Technology (TalTech), Tallinn, Estonia, in 2024. He is currently a researcher at the Centre for Hardware Security at Tallinn University of Technology. 

His research interests primarily revolve around hardware security, with a particular focus on physical design automation and secure ASIC design.
\end{IEEEbiography}

\vskip -1\baselineskip plus -1fil

\begin{IEEEbiography}[{\includegraphics[width=1in,height=1.25in,clip,keepaspectratio]{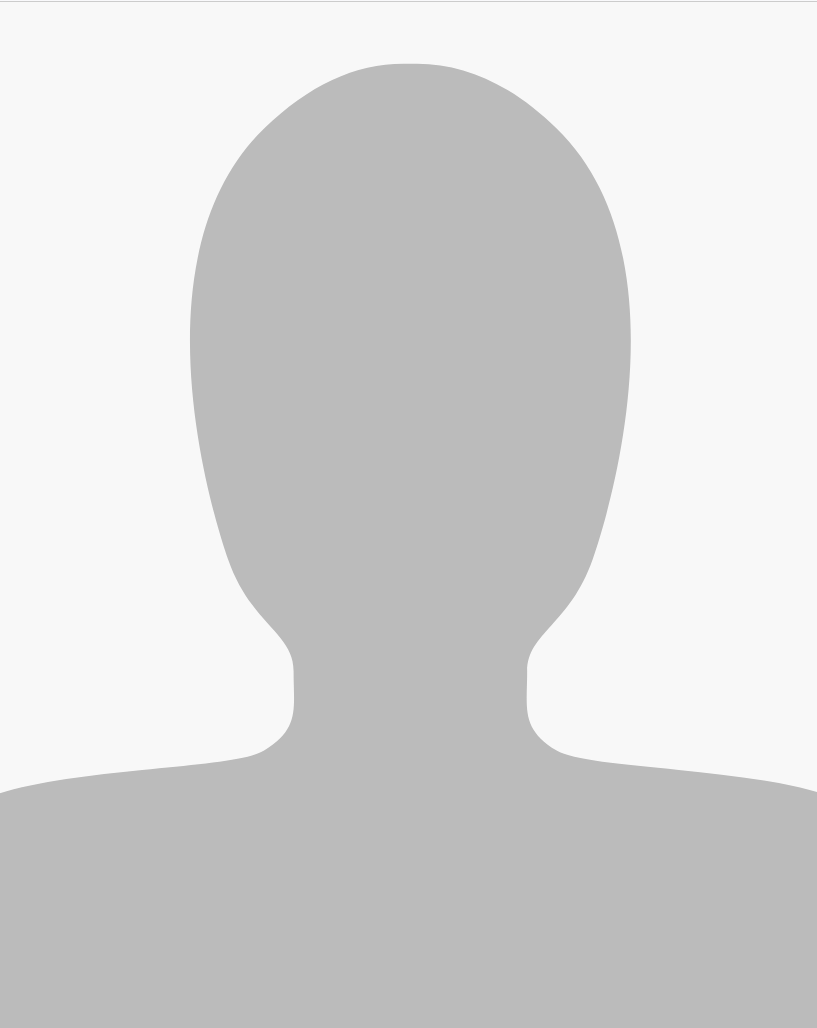}}]{Ashira Johara} is currently pursuing a Bachelor's degree in Electrical and Computer Engineering at Carnegie Mellon University. 

Her research interests are ASIC/VLSI design.
\end{IEEEbiography}

\vskip -1\baselineskip plus -1fil

\begin{IEEEbiography}[{\includegraphics[width=1in,height=1.25in,clip,keepaspectratio]{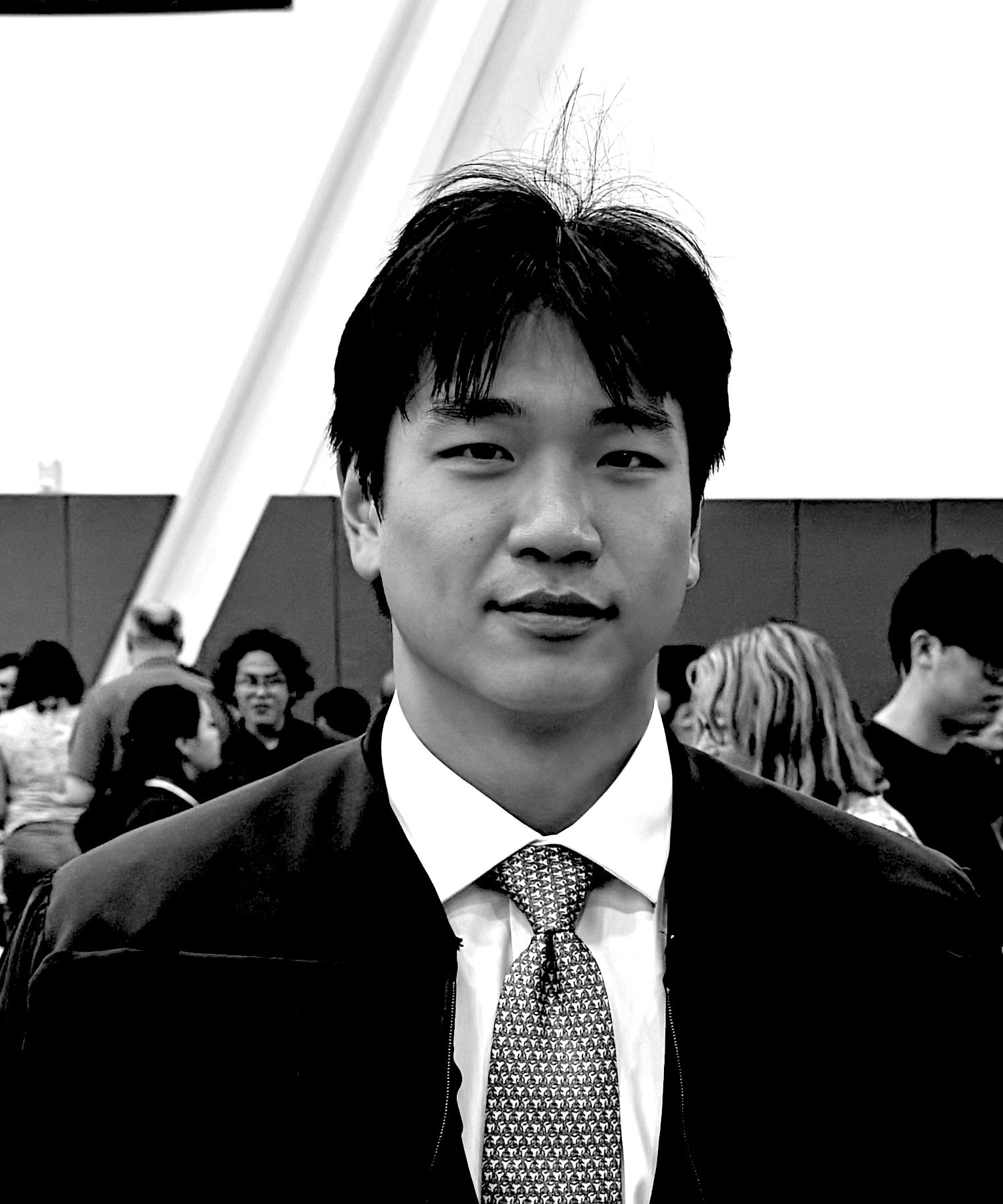}}]{Kyungbin Park} received the Bachelor's degree in Electrical and Computer Engineering at Carnegie Mellon University, Pittsburgh, PA, where he is currently pursuing the Master's degree in the same department. 

His research interests are in silicon engineering, particularly nanoscale lithography and physical design security.
\end{IEEEbiography}

\vskip -1\baselineskip plus -1fil

\begin{IEEEbiography}[{\includegraphics[width=1in,height=1.25in,clip,keepaspectratio]{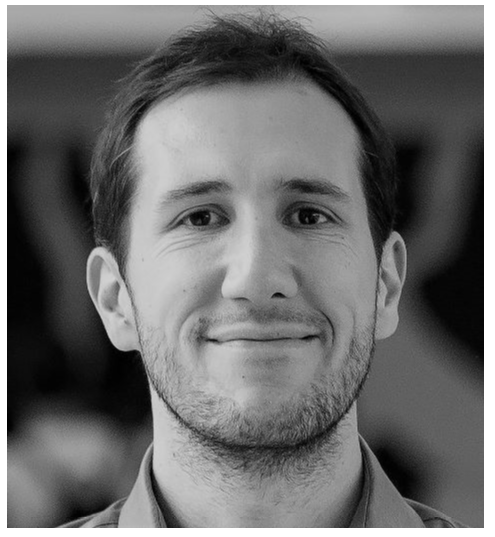}}]{Samuel Pagliarini}
(M'14) received the PhD degree from Telecom ParisTech, Paris, France, in 2013. He has held research positions with the University of Bristol, Bristol, UK, and with Carnegie Mellon University, Pittsburgh, PA, USA. He led the Centre for Hardware Security at Tallinn University of Technology (TalTech) in Tallinn, Estonia, from 2019 to 2024. He is currently a Special Professor at Carnegie Mellon University. His current research interests include many facets of digital circuit design, with a focus on circuit reliability, dependability, and hardware trustworthiness.

\end{IEEEbiography}

\end{document}